\documentclass[twocolumn,english,aps,prl,floatfix,amssymb,superscriptaddress]{revtex4}
\usepackage[latin9]{inputenc}
\usepackage{color}
\definecolor{note_fontcolor}{rgb}{0.800781, 0.800781, 0.800781}
\usepackage{babel}
\usepackage{braket}
\usepackage{amsmath}
\usepackage{amssymb}
\usepackage{graphicx}
\usepackage{esint}
\usepackage{bbold}
\usepackage[unicode=true,pdfusetitle,
 bookmarks=true,bookmarksnumbered=false,bookmarksopen=false,
 breaklinks=false,pdfborder={0 0 0},backref=false,colorlinks=true]
 {hyperref}

\makeatletter

\@ifundefined{textcolor}{}
{%
 \definecolor{BLACK}{gray}{0}
 \definecolor{WHITE}{gray}{1}
 \definecolor{RED}{rgb}{1,0,0}
 \definecolor{GREEN}{rgb}{0,1,0}
 \definecolor{BLUE}{rgb}{0,0,1}
 \definecolor{CYAN}{cmyk}{1,0,0,0}
 \definecolor{MAGENTA}{cmyk}{0,1,0,0}
 \definecolor{YELLOW}{cmyk}{0,0,1,0}
}

\makeatother

\begin{document}

\title{Topological Superconductivity in a Planar Josephson Junction}

\author{Falko Pientka}
\thanks{These authors have contributed equally to this work.}
\affiliation{Department of Physics, Harvard University, Cambridge, Massachusetts 02138, USA}
\author{Anna Keselman}
\thanks{These authors have contributed equally to this work.}
\affiliation{Department of Condensed Matter Physics, Weizmann Institute of Science, Rehovot, Israel 76100}
\author{Erez Berg}
\affiliation{Department of Condensed Matter Physics, Weizmann Institute of Science, Rehovot, Israel 76100}
\author{Amir Yacoby}
\affiliation{Department of Physics, Harvard University, Cambridge, Massachusetts 02138, USA}
\author{Ady Stern}
\affiliation{Department of Condensed Matter Physics, Weizmann Institute of Science, Rehovot, Israel 76100}
\author{Bertrand I. Halperin}
\affiliation{Department of Physics, Harvard University, Cambridge, Massachusetts 02138, USA}

%
\begin{abstract}
We consider a two-dimensional electron gas with strong spin-orbit coupling contacted by two superconducting leads, forming a Josephson junction. We show that in the presence of an in-plane Zeeman field the quasi-one-dimensional region between the two superconductors can support a topological superconducting phase hosting Majorana bound states at its ends. 
We study the phase diagram of the system as a function of the Zeeman field and the phase difference between the two superconductors (treated as an externally controlled parameter). Remarkably, at a phase difference of $\pi$, the topological phase is obtained for almost any value of the Zeeman field and chemical potential. In a setup where the phase is not controlled externally, we find that the system undergoes a first-order topological phase transition when the Zeeman field is varied. At the transition, the phase difference in the ground state changes abruptly from a value close to zero, at which the system is trivial, to a value close to $\pi$, at which the system is topological. The critical current through the junction exhibits a sharp minimum at the critical Zeeman field, and is therefore a natural diagnostic of the transition.
We point out that in presence of a symmetry under a modified mirror reflection followed by time reversal, the system belongs to a higher symmetry class and the phase diagram as a function of the phase difference and the Zeeman field becomes richer.
\end{abstract}
\maketitle

\section{Introduction}

\begin{figure*}

\begin{tabular}{lrr}
\includegraphics[clip=true,trim =0.3cm 0.7cm 0cm 0.0cm,width=0.33\textwidth]{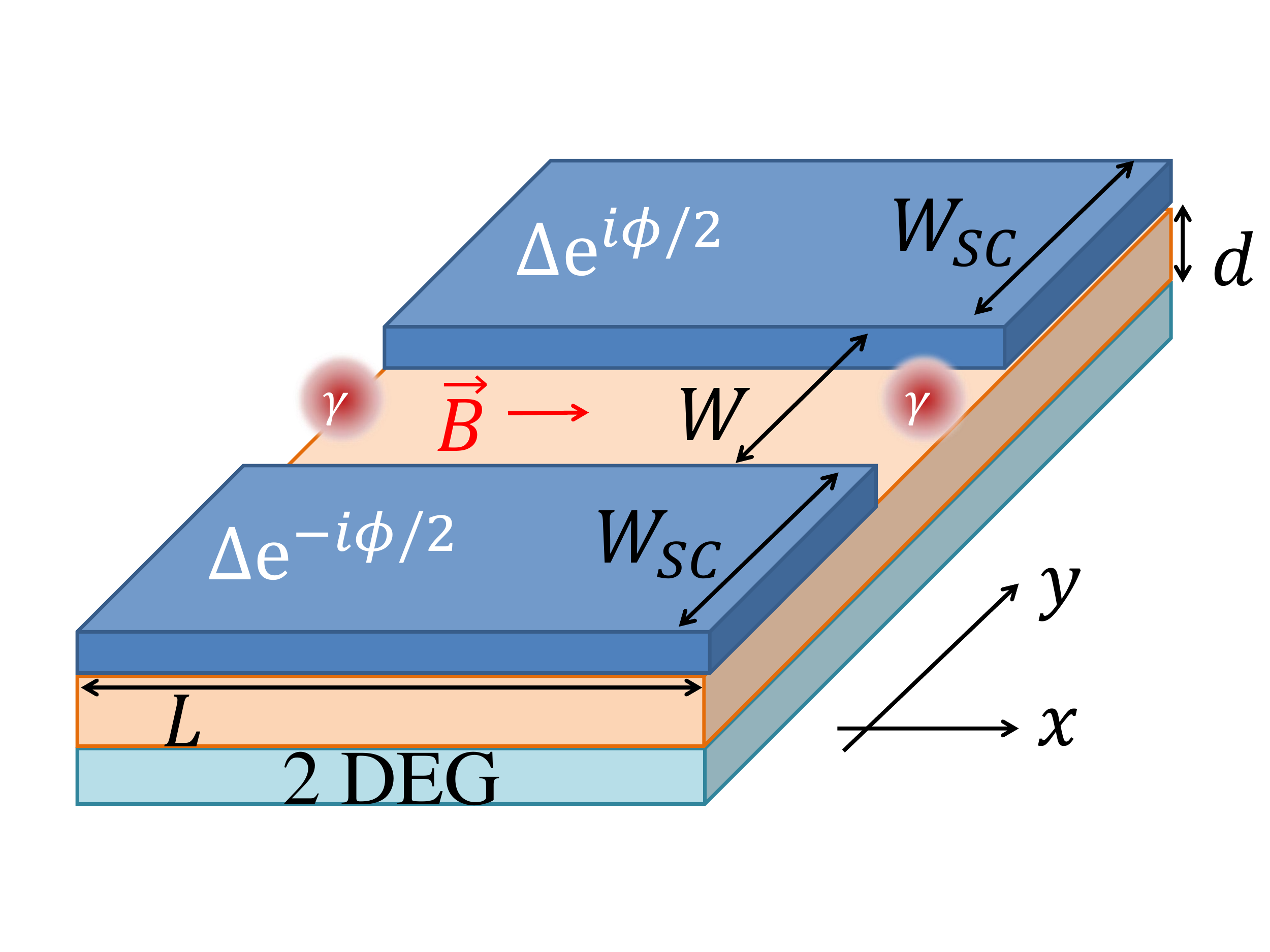}
\llap{\parbox[c]{1.0cm}{\vspace{-7.5cm}\hspace{-10cm}\footnotesize{(a)}}} & \hspace{3mm}
\includegraphics[width=0.28\textwidth]{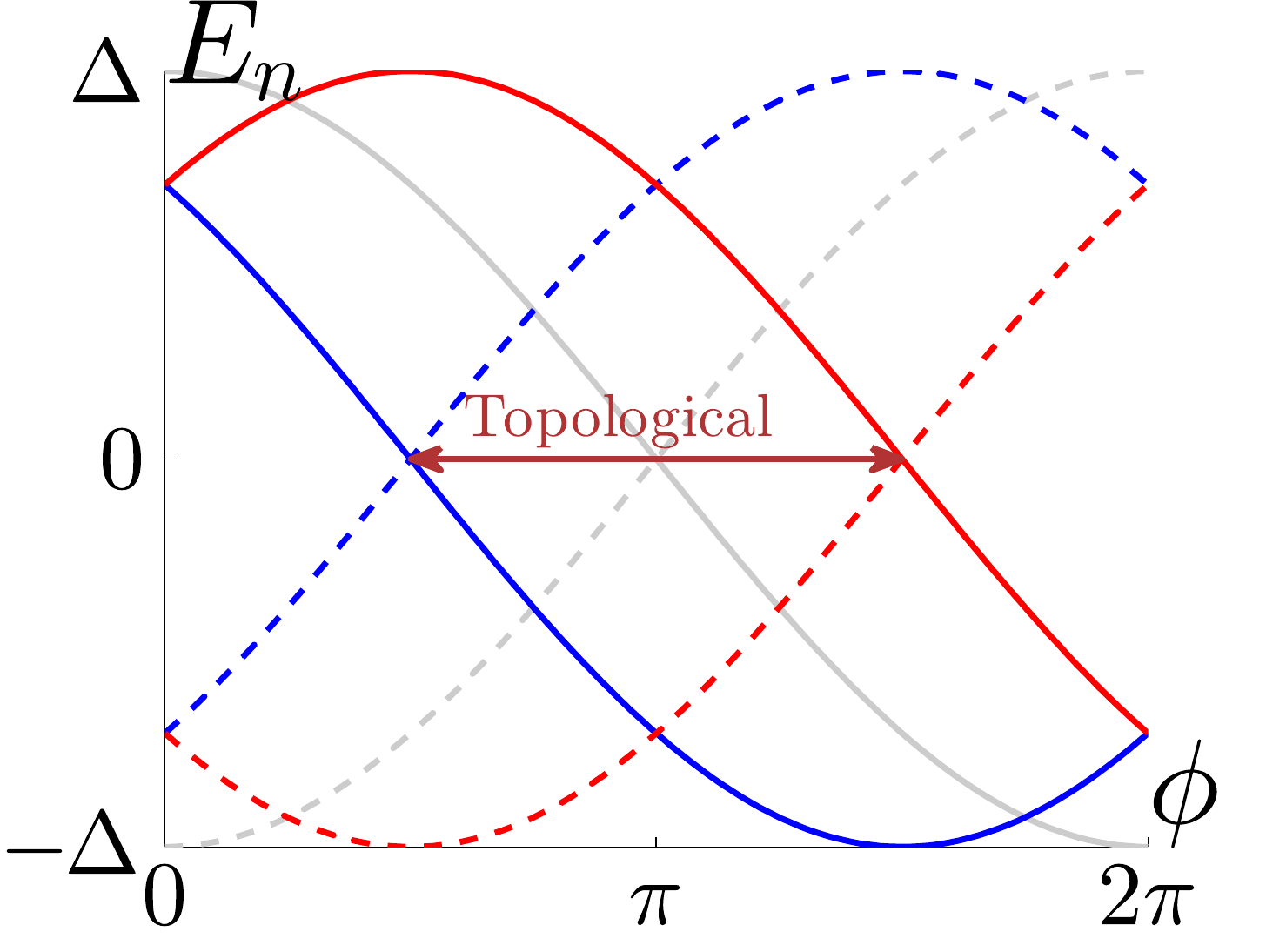}  
\llap{\parbox[c]{0.0cm}{\vspace{-7.5cm}\hspace{-10.2cm}\footnotesize{(b)}}} & \hspace{3mm}
\includegraphics[width=0.28\textwidth]{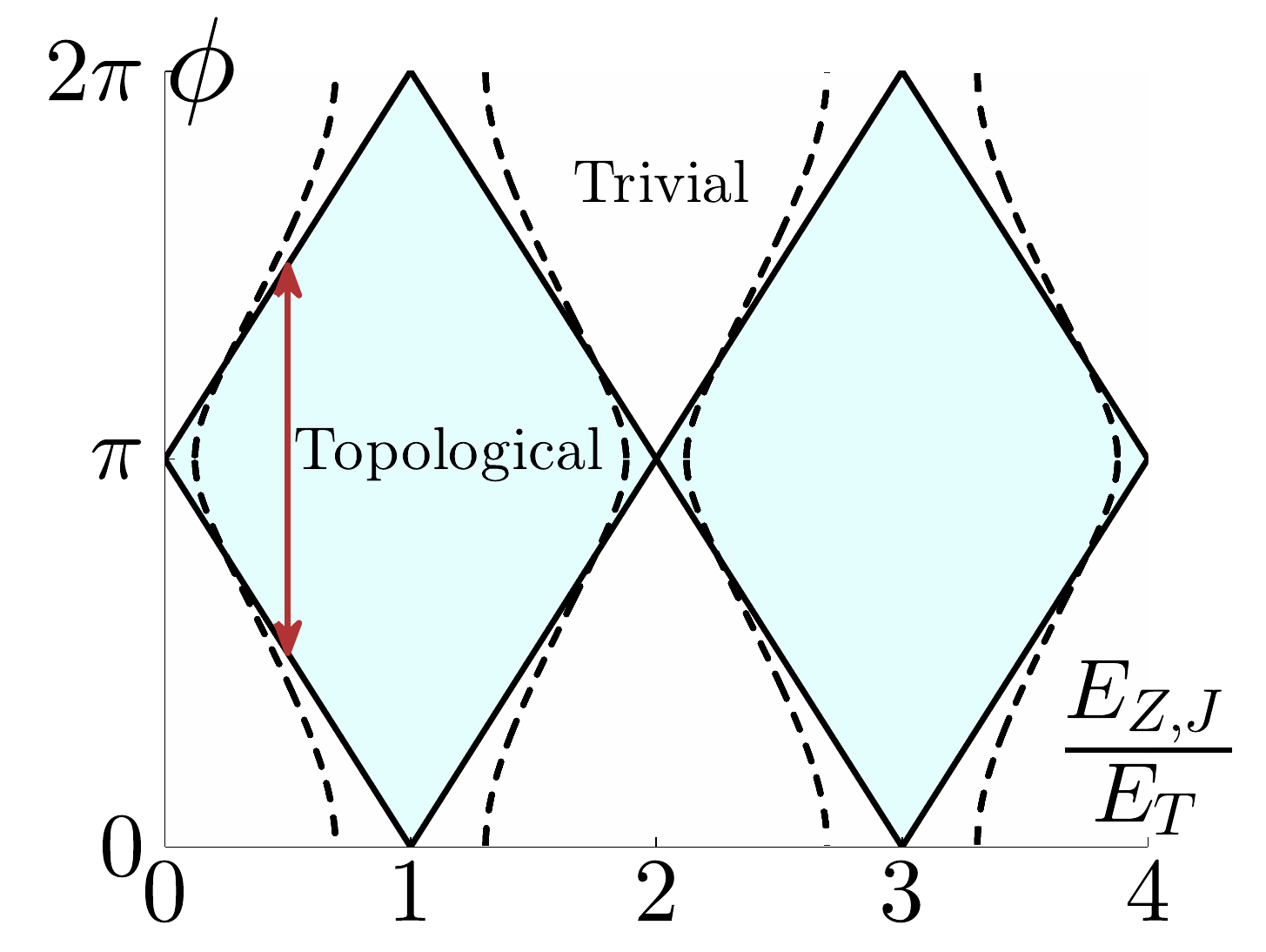}
\llap{\parbox[c]{0.0cm}{\vspace{-7.5cm}\hspace{-10.4cm}\footnotesize{(c)}}}
\end{tabular}

\caption{
(a) A Josephson junction is formed in a 2DEG with Rashba spin-orbit
coupling by proximity coupling it to two s-wave superconductors with
relative phase difference $\phi$. An in-plane magnetic field is applied
parallel to the interface between the normal and the superconducting
regions. 
(b) The bound states spectrum in a narrow junction for $k_x=0$. The spectrum in the absence of a Zeeman field is twofold degenerate and is indicated by the grey lines. In presence of the Zeeman field the spectrum for the two spin states (plotted in red and blue) is split allowing for the appearance of a topological phase. 
(c) Phase diagram as a function of the Zeeman field in the junction,
$E_{{\rm Z,J}}$, given in units of the Thouless energy $E_{{\rm T}}=\left(\pi/2\right)v_{{\rm F}}/W$, and the phase difference $\phi$. The solid lines are the phase boundaries in
the absence of any normal backscattering at the superconducting-normal interface,
while the dashed lines correspond to a junction transparency of $0.75$,
and a phase $k_{{\rm F}}W+\varphi_{N}=3\pi/8$ as defined in Sec.~\ref{sec:Phase-Diagram}. The arrows indicate the range of $\phi$ values between the two zero energy crossings in (b) for which the system is topological.}
\label{fig:setup}
\end{figure*}

Since the realization of two-dimensional topological insulators a decade ago, a plethora of new phases of matter with nontrivial topology in one, two and three dimensions have been discovered in experiment. Considerable experimental and theoretical effort has been dedicated to the study of zero-energy Majorana bound states which arise in topological superconductors as edge states in one dimension or bound to vortices in two dimensions \cite{Alicea2012,Beenakker2013}. Advances in nanotechnology and the prospect of using Majorana states as building blocks of topological quantum computers have triggered intense experimental efforts to realize and characterize them in one-dimensional systems \cite{Mourik2012,Das2012,Nadj-Perge2014,Albrecht2016}. More recently, two-dimensional electron gases (2DEGs) with induced superconductivity \cite{hart2015controlled,Wan2015,Kjaergaard2016} have emerged as a contender for topological superconductivity.

A key challenge for existing one-dimensional platforms such as proximitized semiconductor nanowires \cite{Hyart2013} or atomic chains \cite{Li2016} is to develop networks that allow braiding of multiple Majorana states. An alternative route towards realizing a scalable architecture is to pattern a network of one-dimensional channels into a proximitized 2DEG using gates \cite{hart2015controlled,Shabani2016}. While this approach offers great flexibility in designing networks, it may be cumbersome to drive many channels individually into a topological regime via local gates employing additional local probes. Moreover, gates may change the shape of the channel or physical parameters such as spin-orbit coupling and their effect is strongly influenced by the electrostatic properties of the nearby superconductors \cite{Vuik2016}.

Here we pursue a different strategy to realize Majorana bound states motivated by recent experiments on Josephson junctions in proximity-coupled 2DEGs \cite{hart2015controlled,Wan2015,Kjaergaard2016}. Carriers with energies below the superconducting gap are trapped in the quasi one-dimensional junction region between two superconducting leads as depicted in Fig.~\ref{fig:setup}. In the presence of a Zeeman field, the junction can enter a topological superconducting phase akin to the one in proximitized nanowires and Majorana bound states appear at the ends of the junction.

A key advantage of this setup is that the lateral dimension allows for additional experimental knobs such as a phase difference or a supercurrent across the junction. One of the central results of this work is that a phase bias can induce a robust topological phase in the junction. Most strikingly, in the absence of normal reflection, junctions at a phase difference of $\pi$ host Majorana states to a large extent independently of parameters such as chemical potential, Zeeman field, width of the junction, or induced pairing strength, for as long as the gap in the bulk 2DEG does not close. Moreover, the phase difference can be used as a powerful switch that changes the topology of the entire phase space from trivial at zero to topological at $\pi$. This is in stark contrast to previous proposals which require careful gating and a Zeeman field beyond a critical value.  A setup based on Josephson junctions may also facilitate the realization of topological superconductor networks. By tuning a global phase difference multiple Josephson junctions can be tuned simultaneously into a topological phase without tuning local parameters or requiring local probes.

In the presence of normal reflection in the junction or at the interface to the superconductor deviations from this ideal behavior occur. As long as normal reflection is not too strong, however, our results still hold in extended regions of the parameter space. 

On the face of it, the system we consider belongs to class D in the ten-fold classification \cite{AltlandZirnbauer1997}, since time-reversal symmetry is broken and particle-hole symmetry holds. 
In fact, our system has an additional symmetry given by a combination of a mirror reflection, time-reversal, and a gauge transformation, which places it in class BDI (see also Ref. \cite{Mizushima2013}). Interestingly, this symmetry is present for \emph{any} value of the phase difference between the superconductors.
As a consequence, slivers with additional topological phases appear in the phase diagram as a function of the in-plane Zeeman field and the phase difference between the superconductors. The system is brought back to class D if the magnitude of the superconducting gap on the two sides of the junction is different.

If the phase difference is not imposed externally the system can undergo a first order phase transition in which the phase jumps from a phase close to $0$ to a phase close to $\pi$ with increasing in-plane magnetic field. Similar transitions have previously been studied in ferromagnetic Josephson junctions \cite{hart2015controlled,Ryazanov2001,Kontos2002,Frolov2004}. 
Quite remarkably, our results suggest that such a first order phase transition in the present setup is in fact a topological phase transition unique to the two dimensional geometry. The system can thus self tune into a topological phase when the magnetic field is varied and realizes a first-order topological phase transition without gap closing. 
Moreover, this transition is accompanied by a minimum of the critical current. Therefore, the critical current can serve as an inherent probe of the topological phase transition. Surprisingly, the contrast of the critical current modulation with the field increases with temperature. At high temperatures the critical current vanishes at the magnetic field of the underlying zero-temperature topological transition.
This insight suggests that the experimental results presented by Hart {\it et al.} \cite{hart2015controlled} indicate an underlying topological phase transition in the ground state.

This paper is organized as follows. We start by presenting
the proposed setup and a summary of our results in Sec.~\ref{sec:Summary}.
We then show the derivation of the phase diagram for the system as
function of the phase difference and the Zeeman field, and discuss
the magnitude of the topological gap and the appearance of Majorana end modes in Sec.~\ref{sec:Phase-Diagram-and-Gap}.
In Sec.~\ref{sec:critical_current} we discuss the first order topological phase transition as function of the Zeeman field and how the critical current can serve as a novel experimental probe to detect this transition in the suggested setup. We conclude with discussion of the presented results in Sec.~\ref{sec:Discussion}. The paper is followed by appendices
that cover several technical details.

\section{Physical picture and summary of results\label{sec:Summary}}

\begin{figure}[!ht]
\includegraphics[width=0.3\textwidth]{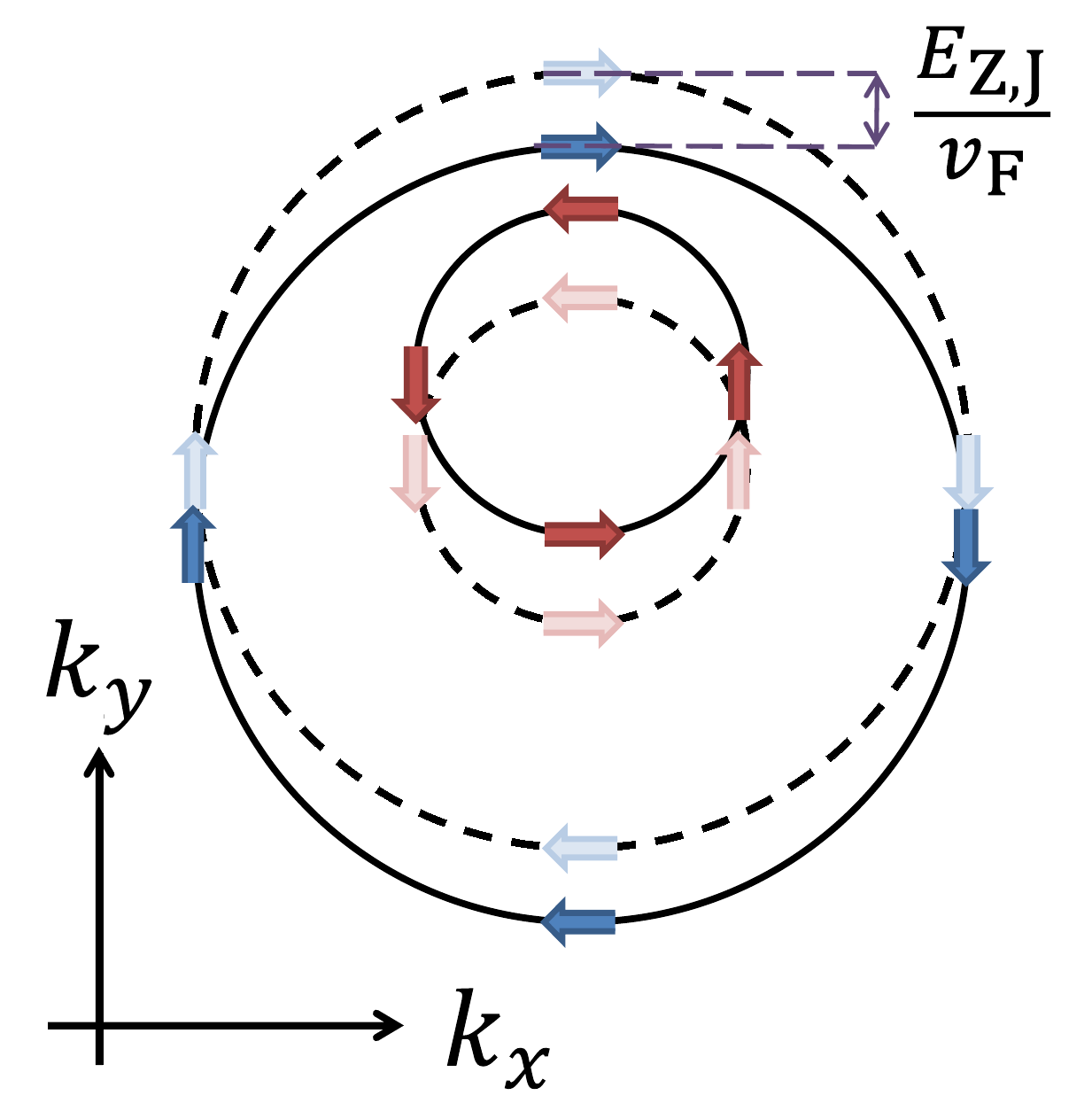}
\llap{\parbox[c]{1.0cm}{\vspace{-10cm}\hspace{-12.5cm}\footnotesize{(a)}}} 

\includegraphics[clip=true,trim =0cm 2.5cm 0cm 0cm,width=0.47\textwidth]{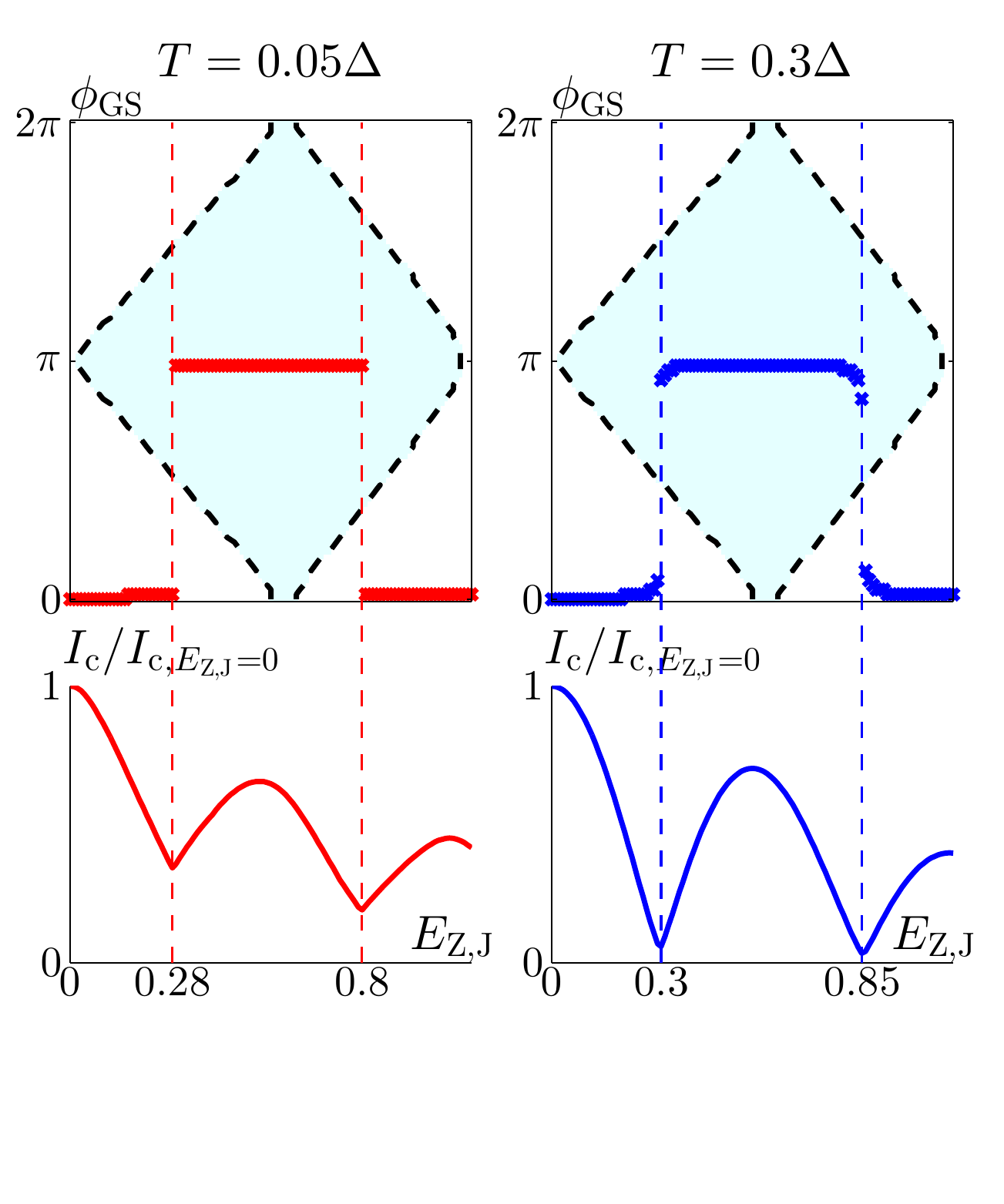}
\llap{\parbox[c]{0.0cm}{\vspace{-16.5cm}\hspace{-16.5cm}\footnotesize{(b)}}}
\caption{(a) A Zeeman field along $x$ shifts the two Rashba-split Fermi surfaces of the 2DEG in opposite directions along $y$. The arrows indicate the orientation of the spin at each point on the Fermi surface. 
(b) The phase difference, $\phi_{\rm GS}$, that minimizes the ground state energy (upper panel) and the critical current modulation (lower panel) as function of the Zeeman field obtained numerically using a tight binding model for the system  \cite{TBParams}. 
Left (right) panel corresponds to a temperature of $T=0.05\Delta$ ($T=0.3\Delta$). 
(Note that we set $k_{\rm B}=1$ throughout the manuscript.) The light blue color indicates the region in the parameter space for which the system is in the topological phase.
As the Zeeman field is varied, the system undergoes a series of first order topological phase transitions, in which $\phi_{\rm GS}$  changes abruptly between values lying in the topological and trivial regions of the phase diagram. The critical current exhibits minima at the points of the phase transitions. As the temperature is increased the minima become deeper.}
\label{fig:Ic}
\end{figure}

We consider a two-dimensional semiconductor with Rashba spin-orbit coupling, partially covered with two superconducting contacts in an in-plane magnetic field as depicted in Fig.~\hyperref[fig:setup]{\ref{fig:setup}(a)}. For the most part, we will be interested in the case of an infinite system, where the width of the leads and the length of the junction $W_{SC},L\to \infty$, while the separation of the leads, $W$, remains finite. We describe the system by a Bogoliubov--de Gennes Hamiltonian in the Nambu basis $(\psi_{\uparrow}^{\phantom{}},\psi_{\downarrow}^{\phantom{}},\psi_{\downarrow}^{\dag},-\psi_{\uparrow}^{\dag})$
\begin{align}
H= & \left(\frac{k_{x}^{2}-\partial_{y}^{2}}{2m}-\mu+\frac{m\alpha^{2}}{2}\right)\tau_{z}+\alpha(k_{x}\sigma_{y}+i\partial_{y}\sigma_{x})\tau_{z}\nonumber \\
 & +E_{{\rm Z}}(y)\sigma_{x}+\Delta(y)\tau_{+}+\Delta^*(y)\tau_{-}.\label{hamil}
\end{align}
Here $k_x$ is the momentum along $x$ which is conserved in the system (we set $\hbar=1$ throughout the manuscript), $m$ is the effective mass of the 2DEG, $\mu$ is the chemical potential measured from the bottom of the spin-orbit split bands, $\alpha$ is the strength of Rashba spin-orbit coupling and $E_{{\rm Z}}(y)=g(y)\mu_{B}B/2$ is the Zeeman energy induced by an external magnetic field. 
We assume different $g$ factors in the junction and underneath the superconducting leads and denote 
\begin{align}
E_{{\rm Z}}(y)=E_{{\rm Z,L}}\theta(|y|-W/2)+E_{{\rm Z,J}}\theta(W/2-|y|),
\end{align} 
where $\theta(x)$ is a step function. For simplicity we focus on the case of zero Zeeman field underneath the leads and postpone the discussion of nonzero $E_{{\rm Z,L}}$ to Sec.~\ref{sec:Phase-Diagram}.
The proximity induced pairing in the semiconductor 2DEG is accounted for by 
\begin{align}
\Delta(y)=\Delta e^{i{\rm sgn}(y)\phi/2}\theta(|y|-W/2),
\end{align}
where $\phi$ is the phase difference between the two superconductors. The Pauli matrices $\sigma,\tau$ act in the spin and particle-hole basis respectively, and $\tau_\pm=\tau_x\pm i\tau_y$.

States at subgap energies are confined to the quasi-one-dimensional junction between the two superconducting leads. Under suitable conditions the junction can enter a one-dimensional (1d) topological superconducting phase. 
In this paper we study two experimental configurations in which the model described by Eq.~\eqref{hamil} and Fig.~\hyperref[fig:setup]{\ref{fig:setup}(a)} may be realized. In the first configuration the phase across the junction is a parameter controlled externally by applying a current or a magnetic flux. In the second configuration the phase is left to self-tune so as to minimize the ground state energy.
For the first configuration we calculate the phase diagram as a function of the phase across the junction and the Zeeman field, while for the second configuration we identify the conditions under which the system self tunes to a topological phase. We find that the critical current of the junction can be used as a probe for the transitions between topological and trivial phases.

We start by evaluating the topological index for particle-hole symmetric systems in class D. 
As we shall see in Sec.~\ref{sec:BDI}, the model has a higher symmetry involving a mirror reflection followed by time reversal that places it into the BDI class. Each topological (trivial) region in the phase diagram of class D will be split into subregions with an odd (even) $\mathbb{Z}$ invariant \cite{Kitaev2009,Schnyder2009,foot1}.\nocite{Fidkowski2010} Breaking this symmetry stabilizes the topological phase with a single Majorana bound state at each end. 

The index in class D can be defined as the fermion parity of the ground state at $k_x=0$ \cite{Kitaev2001}. Therefore a phase transition between the trivial and topological classes must be accompanied by a single gap closing at $k_x=0$. At this momentum spin along the $x$ direction is conserved by the Hamiltonian in Eq.~\eqref{hamil} and the spin-orbit coupling can be gauged away by substituting $\partial_{y}\to \partial_{y}+im\alpha\sigma_x$. We arrive at the effective Hamiltonian
\begin{align}
H_0=(-\partial_y^2/2m-\mu)\tau_z+E_{{\rm Z}}\left(y\right)\sigma_x+\Delta\left(y\right)\tau_++\Delta\left(y\right)^*\tau_-\label{hamil_1d} 
\end{align}
describing a one-dimensional Josephson junction in a magnetic field. Figure \ref{fig:setup}(b) shows the subgap spectrum of a narrow junction in the Andreev limit $\mu\gg \Delta$, where normal reflection is absent. At vanishing Zeeman field, the spectrum is twofold degenerate and the system is trivial for all values of $\phi$. At nonzero fields the degeneracy is split opening a topological phase around $\phi=\pi$. At the zero-energy crossings the fermion parity changes and the junction undergoes topological phase transitions.

The resulting phase diagram as a function of Zeeman field and phase bias is shown in Fig.~\hyperref[fig:setup]{\ref{fig:setup}(c)}. Most strikingly the junction is in a topological phase at $\phi=\pi$ for arbitrary Zeeman fields except at isolated values given by even integer multiples of the ballistic Thouless energy of the junction $E_{{\rm T}}=(\pi/2)v_{{\rm F}}/W$. In contrast, at zero phase difference the system remains trivial throughout. As will be shown in Sec.~\ref{sec:Phase-Diagram}, this result generalizes to junctions of arbitrary width as long as normal reflection can be neglected and the system remains gapped. The $\mathbb{Z}_2$ topological index cannot change at $\phi=0,\pi$ because the spectrum is always doubly degenerate at $k_x=0$ and topological phase transitions thus come in pairs. Hence, an externally applied phase bias is a powerful experimental knob, that allows one to tune the topology of the junction to a large extent independent of microscopic parameters.

We can qualitatively understand the effect of weak normal reflection on the phase diagram from the subgap spectra shown in Fig.~\hyperref[fig:setup]{\ref{fig:setup}(b)}. Normal backscattering couples left and right movers, lifting the degeneracy of Andreev levels at $\phi=0,\pi$. Hence the system becomes topological (trivial) in a small range of Zeeman fields at $\phi=0$ ($\phi=\pi$), respectively. The avoided level crossings translate to avoided crossings of phase transition lines as indicated by the dashed lines in Fig.~\hyperref[fig:setup]{\ref{fig:setup}(c)}. As long as normal reflection is not too strong, it remains possible to induce a topological phase by a phase bias in extended regions of parameter space.

We next consider the second configuration in which the phase is determined by the condition that the ground state energy is minimal. Remarkably, we shall see in Sec.~\ref{sec:critical_current} that in this case the system self tunes to the topological phase in a broad range of Zeeman fields, exhibiting a first order topological phase transition. Such a transition will be accompanied by an abrupt change in various thermodynamic quantities characterizing the system, e.g. the magnetization, as well as in the energy gap in the bulk.

The origin of the first order transition is that the phase difference, $\phi_{\rm GS}$, that minimizes the ground state energy changes abruptly between two distinct values, one in the trivial region and one in the topological region, at certain values of the Zeeman field. As a consequence, the junction is expected to show a hysteretic behavior as the Zeeman field is swept at low temperatures. Moreover, we find that the critical current exhibits a minimum at these values of the Zeeman field as shown in Fig.~\hyperref[fig:Ic]{\ref{fig:Ic}(b)}. The critical current can thus be used as a novel experimental probe of the topological phase transitions in this conguration.

These findings can be understood semiclassically in the limit $E_{{\rm Z,J}}\ll\alpha k_{{\rm F}}\ll\mu$.
Due to the Rashba-induced spin-momentum locking, the Zeeman field shifts the two Fermi surfaces uniformly along $k_y$ in opposite directions as depicted in Fig.~\hyperref[fig:Ic]{\ref{fig:Ic}(a)}. This induces a nonzero center of mass momentum $q=2E_{{\rm Z,J}}/v_{{\rm F}}$ in Cooper pairs traversing the junction. Thus the wavefunction of a Cooper pair leaving one superconducting lead can be described by a linear combination of singlet and triplet contributions $\cos(qy)\ket{S}+\sin(qy)\ket{T}$. 
For $qW>\pi/2$ (or, equivalently, $E_{{\rm Z,J}}>E_{\rm T}/2$) the singlet wavefunction has opposite signs at the two superconducting leads and $\phi_{\rm GS}$ switches from $0$ to $\pi$. 
As discussed above, the system is trivial at $\phi=0$ and topological at $\phi=\pi$ in a wide range of parameters. We see therefore, that for $E_{{\rm Z,J}}>E_{{\rm T}}/2$ the system self-tunes into a topological phase via a first order phase transition.
Moreover, at the $0-\pi$ transition point ($E_{{\rm Z,J}}=E_{{\rm T}}/2$) the singlet component, which carries the supercurrent, has a node at the second interface resulting in a vanishing critical current. 
Beyond the semiclassical approximation, we find that the critical current remains nonzero but assumes a local minimum at the transition, as shown by the numerical resuts in Fig.~\hyperref[fig:Ic]{\ref{fig:Ic}(b)}.

When the constraint $\alpha k_{{\rm F}}\ll\mu$ is lifted, the phase difference in the ground state is not necessarily $0$ or $\pi$ and varies with $E_{\rm Z,J}$ \cite{Dolcini2016}.
However, generically, the system still exhibits a jump in $\phi_{\rm GS}$ as a function of Zeeman field. This jump is accompanied by a change of fermion parity at $k_x=0$ and, therefore, it coincides with a topological phase transition even in the more general case.
Moreover, the critical current still exhibits a minimum at the phase transition point, $E_{{\rm Z,J}}=E_{{\rm T}}/2$.
Surprisingly, the minimum grows sharper with increasing temperature, as can be seen in Fig.~\hyperref[fig:Ic]{\ref{fig:Ic}(b)}.

It is encouraging that the modulation of the critical current as function of an in-plane magnetic field, and in particular its revival, has been observed in experiment realizing the setup we consider \cite{hart2015controlled}, indicating that the topological regime in Josephson junctions is within reach of current experiments even in the absence of a phase bias.
Moreover, our theoretical analysis strongly suggests that the vanishing of the critical current as a function of magnetic field, observed in Ref. \cite{hart2015controlled}, is indicative of a first-order topological phase transition.

A direct signature of topological superconductivity can be obtained by a straightforward extension of the setup in Fig.~\hyperref[fig:setup]{\ref{fig:setup}(a)}, when the system has a large but finite length $L$.
Adding a tunnel probe at one end of the junction would enable the detection of Majorana bound states via tunneling conductance measurements. While the conductance should exhibit a zero-bias peak at the end of the junction, no such feature is expected when tunneling into the center of the junction.

\section{The $E_Z-\phi$ Phase Diagram, Topological Gap, and Majorana end modes \label{sec:Phase-Diagram-and-Gap}}

\subsection{Class D Phase Diagram}\label{sec:Phase-Diagram}

As was discussed in the previous section, topological phase transitions that change the parity of the number of Majorana end modes occur when there are zero-energy solutions of the model at $k_{x}=0$, given by Eq.~\eqref{hamil_1d}.
We use scattering theory to obtain the bound state spectrum of the system and in particular to find the conditions for the formation of a zero-energy state.
We work in the limit $\mu\gg\Delta$ and assume at first that there is no normal reflection at the superconducting-normal interfaces.
In this case, the eigenstates decompose into those with left and right-moving currents.  We denote the junction's transmission amplitude for electrons (holes) by $t_{{\rm e\left(h\right)}}$ and the Andreev reflection amplitudes by $r_{{\rm A}}^{\pm}=\exp(i\eta\pm i\phi/2)$,
where $\eta=\cos^{-1}\left[(E-E_{\rm Z,L})/\Delta\right]$, and the sign corresponds to the current direction \footnote{Note that left-moving electrons and right-moving holes acquire the
same amplitude $r_{{\rm A}}^{+}$ upon Andreev reflection.}. In the limit $\mu\gg E_{{\rm Z,J}}$ we can approximate 
\begin{equation}
t_{{\rm e\left(h\right)}}=\exp\left[\pm ik_{{\rm F}}W + i\frac{\left(E-E_{{\rm Z,J}}\right)}{v_{{\rm F}}}W\right],\label{eq:t_e,h}
\end{equation}
where $k_{{\rm F}}=(2m\mu)^{1/2}$ and $v_{{\rm F}}=k_{{\rm F}}/m$
are the Fermi momentum and velocity respectively. The bound state
spectrum can be obtained from the condition $1=\left(r_{{\rm A}}^{\pm}\right)^{2}t_{{\rm e}}t_{{\rm h}}$
\cite{Beenakker1991}.

We arrive at the following condition for the subgap spectrum
\begin{equation}
\cos^{-1}\left(\frac{E_{n}-E_{\rm Z,L}}{\Delta}\right)=\frac{\pi}{2}\frac{E_{n}}{E_{{\rm T}}}-\frac{\pi}{2}\frac{E_{{\rm Z,J}}}{E_{{\rm T}}}\pm\frac{\phi}{2}+n\pi,\ n\in\mathbb{Z}.\label{eq:bound_states_cond}
\end{equation}
This equation implies a twofold degeneracy of the spectrum at $\phi=0$
and $\phi=\pi$. This degeneracy is a consequence of a mirror symmetry
and the absence of normal reflection from the superconducting leads.
As an important consequence
of this degeneracy the $\mathbb{Z}_{2}$ topological index cannot
change at $\phi=0,\pi$ as zero-energy crossings always come in pairs.
We first consider the case $E_{\rm Z,L}=0$. Equation \eqref{eq:bound_states_cond} then has zero-energy solutions for
\begin{equation}
\frac{\pi}{2}\frac{E_{{\rm Z,J}}}{E_{{\rm T}}}\pm\frac{\phi}{2}=\frac{\pi}{2}+\pi n.\label{eq:phase_boundar_vanishing_B}
\end{equation}
This condition sets the phase boundaries for the phase diagram. It creates a diamond structure with alternating trivial and topological regions as can be seen in Fig.~\hyperref[fig:setup]{\ref{fig:setup}(c)}.

\begin{figure}
\begin{tabular}{cc}
\includegraphics[width=0.235\textwidth]{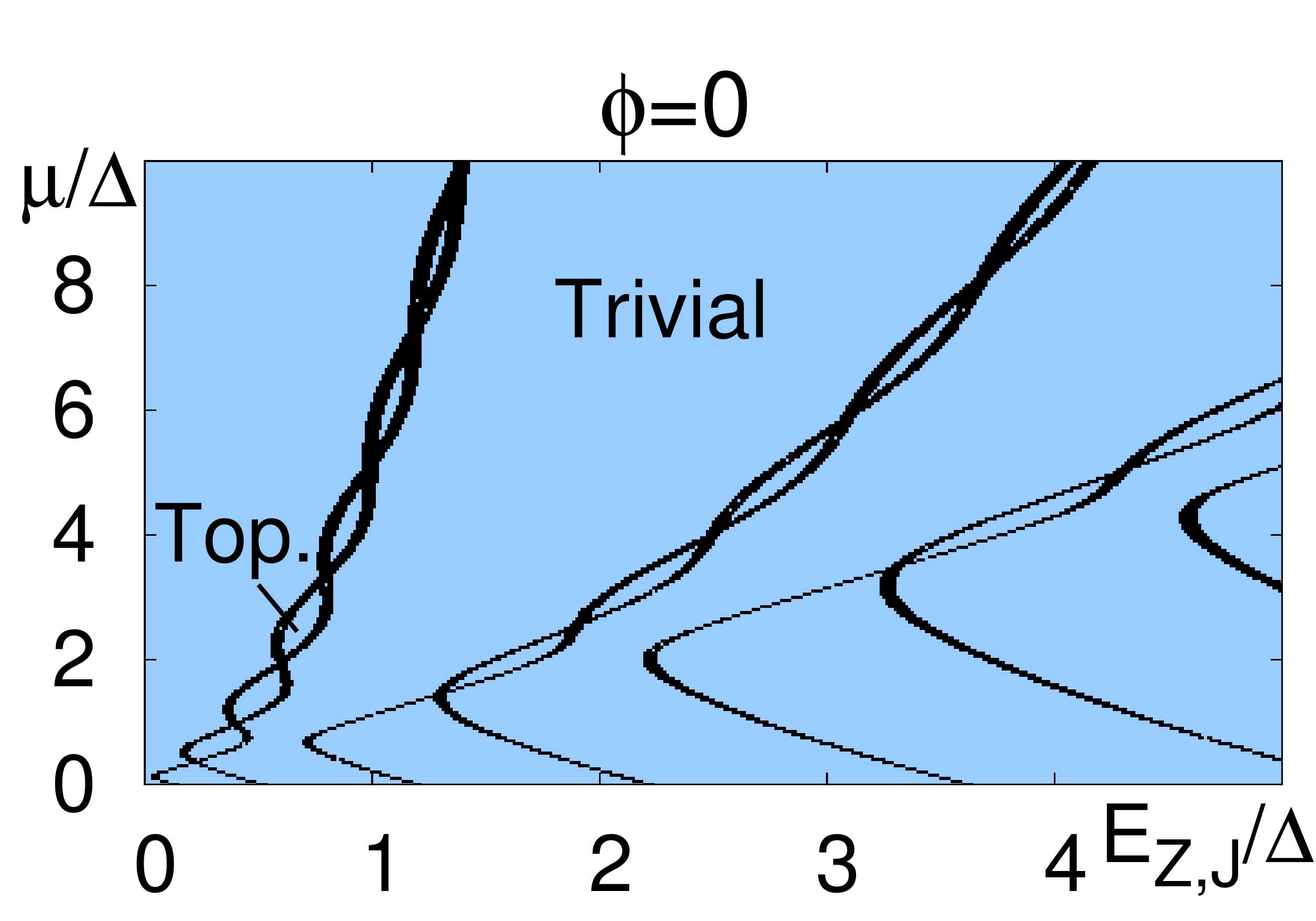}  
\llap{\parbox[c]{0.0cm}{\vspace{-5.3cm}\hspace{-8.6cm}\footnotesize{(a)}}} & 
\includegraphics[width=0.235\textwidth]{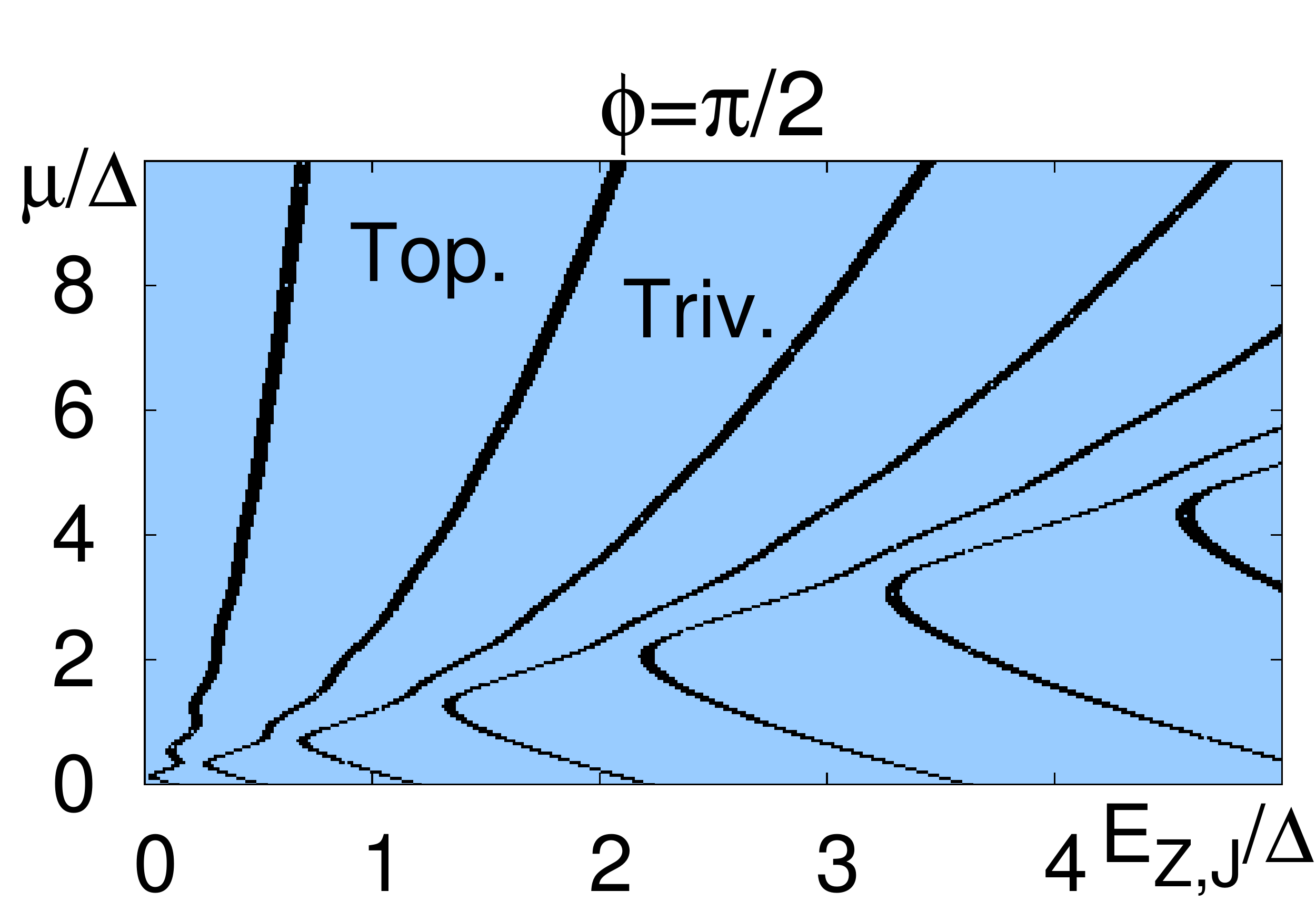}  
\llap{\parbox[c]{0.0cm}{\vspace{-5.3cm}\hspace{-8.6cm}\footnotesize{(b)}}} \\
\includegraphics[width=0.235\textwidth]{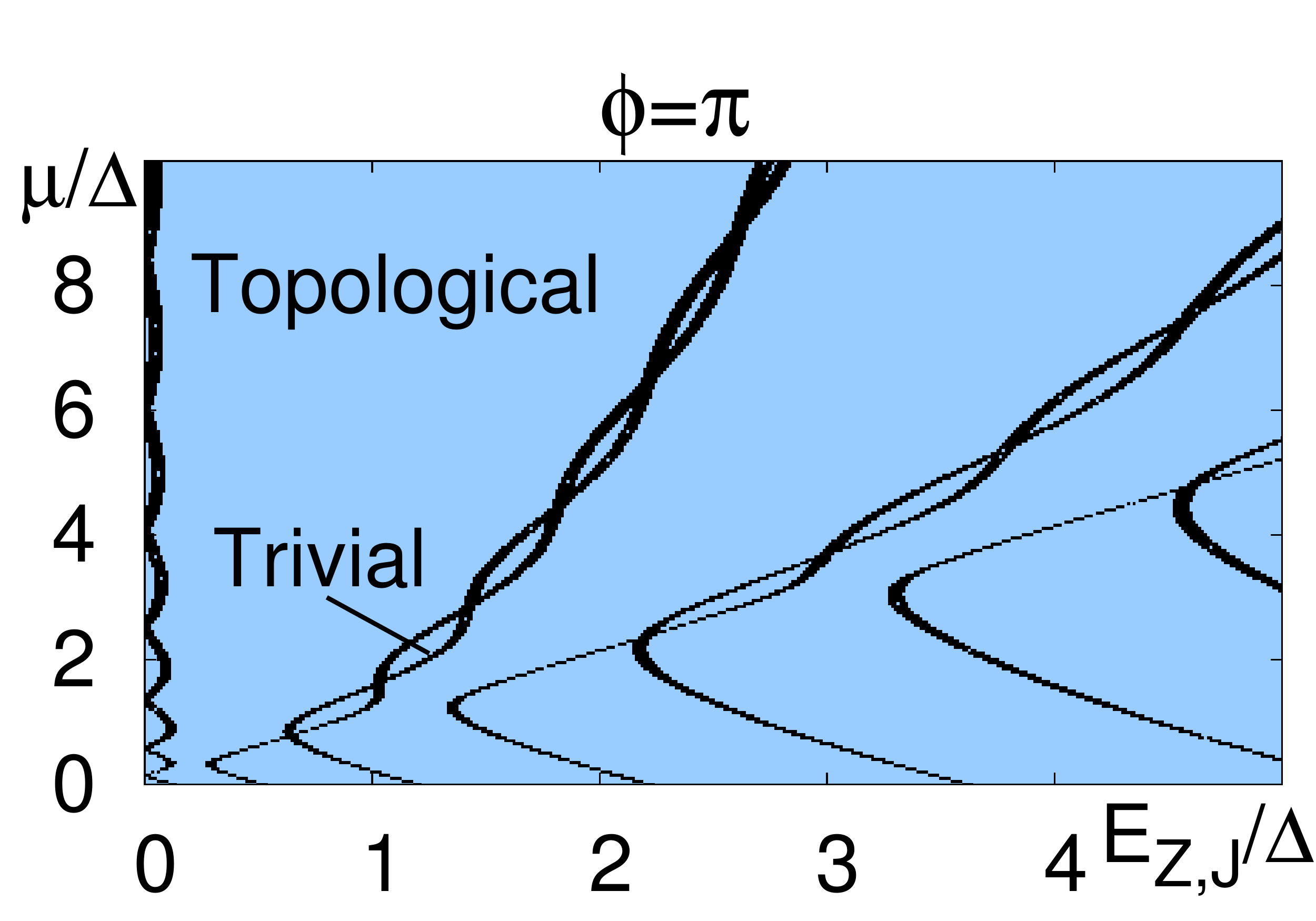}
\llap{\parbox[c]{0.0cm}{\vspace{-5.3cm}\hspace{-8.6cm}\footnotesize{(c)}}} &
\includegraphics[width=0.235\textwidth]{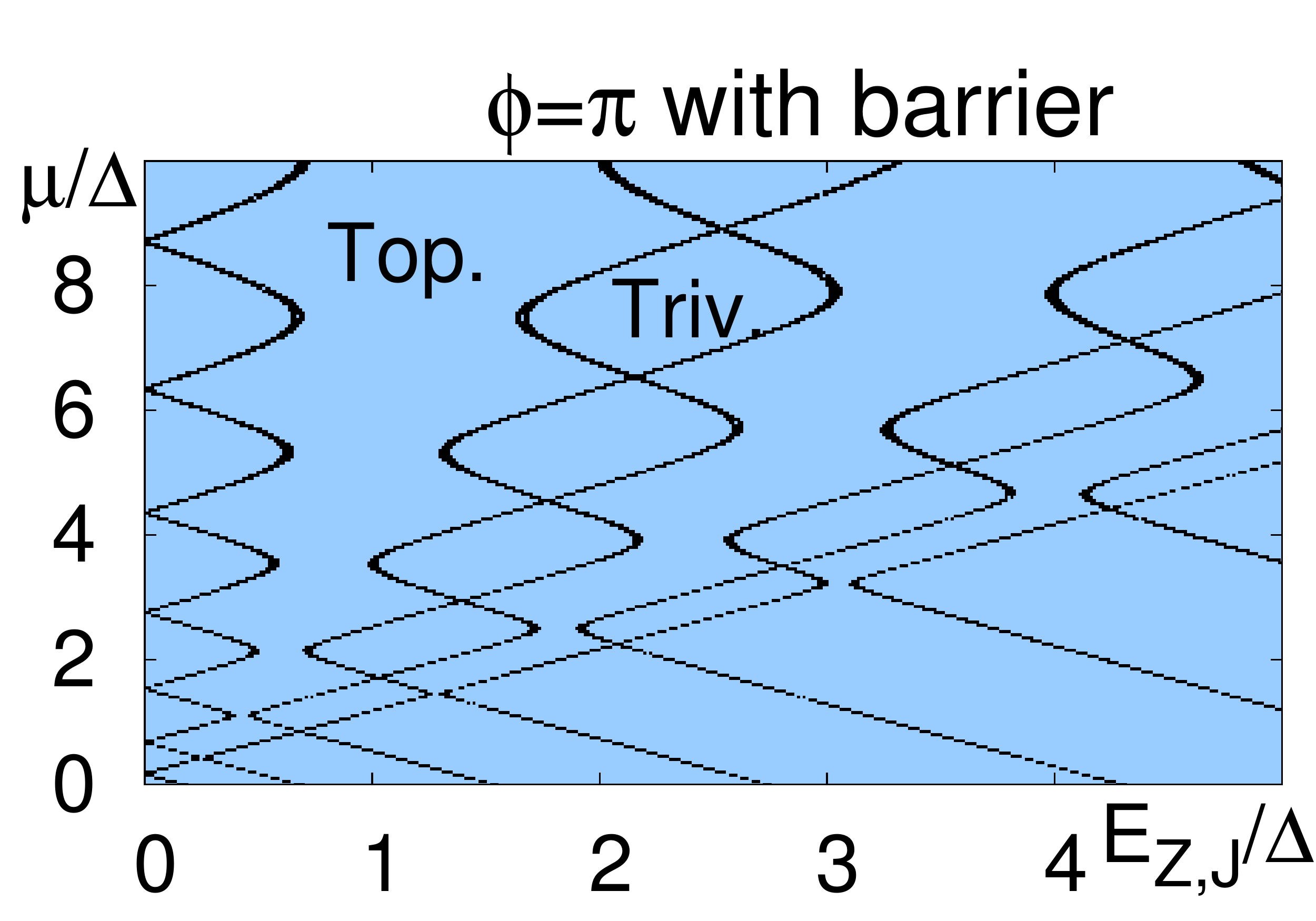}
\llap{\parbox[c]{0.0cm}{\vspace{-5.3cm}\hspace{-8.6cm}\footnotesize{(d)}}}

\end{tabular}
\caption{(a-c) Numerical results for the phase diagram as function of the Zeeman
field and the chemical potential for different values of $\phi$,
obtained using the scattering matrix approach (see Appx.~\ref{app:ScatteringMatrix} for details). The width of the junction is $W=5(m\Delta)^{-1/2}$. The phase boundaries
are determined by a gap closing at $k_{x}=0$. At low chemical potentials
$\mu\simeq\Delta$ a finite amount of normal reflection from the superconducting
leads results in oscillatory modulations of the phase boundaries as function of $\mu$.
The topological phase significantly expands as the phase difference
between the leads is tuned from 0 to $\pi$. (d) Same as (c) but with a rectangular potential barrier at each interface between lead and junction. The barrier has height $200\Delta$ and width $0.01(m\Delta)^{-1/2}$. This corresponds to a transparency of 0.82 at $\mu=10\Delta$. The presence of a reflecting barrier leads to stronger modulations of the boundaries.}
\label{fig:PD_Bmu_num} 
\end{figure}

For nonvanishing normal reflection probability at the superconducting-normal interface the equation for the bound states is identical to Eq. \eqref{eq:bound_states_cond} with $\phi$ replaced by 
\begin{align}
\tilde{\phi} & =\cos^{-1}\left[r^{2}\cos(2k_{{\rm F}}W+2\varphi_{N})+(1-r^{2})\cos\phi\right].\label{eq:phi_tilde}
\end{align}
The phase $\varphi_{N}$ is defined in Eq.~\eqref{eq:reflection_amplitudes} and depends on the details of the normal reflection. (See Appx.~\ref{app:normal_reflection} for the derivation of this result). 
The phase boundaries are therefore given by Eq.~\eqref{eq:phase_boundar_vanishing_B},
with $\phi\to\tilde{\phi}$. These are depicted as dashed lines in
Fig.~\hyperref[fig:setup]{\ref{fig:setup}(c)}. As expected, the degeneracy at $\phi=0$
and $\phi=\pi$ is removed, and the topological (trivial) phase can
now be obtained for some range of Zeeman fields at $\phi=0$ ($\phi=\pi$).
Due to the dependence of $\tilde{\phi}$ on $k_{{\rm F}}$,  finite
reflection amplitude will result in oscillatory modulations of the
phase boundaries as function of $\mu$ as we show in Fig. \ref{fig:PD_Bmu_num}.

\begin{figure}[tb]
\centering
\includegraphics[width=0.4\textwidth]{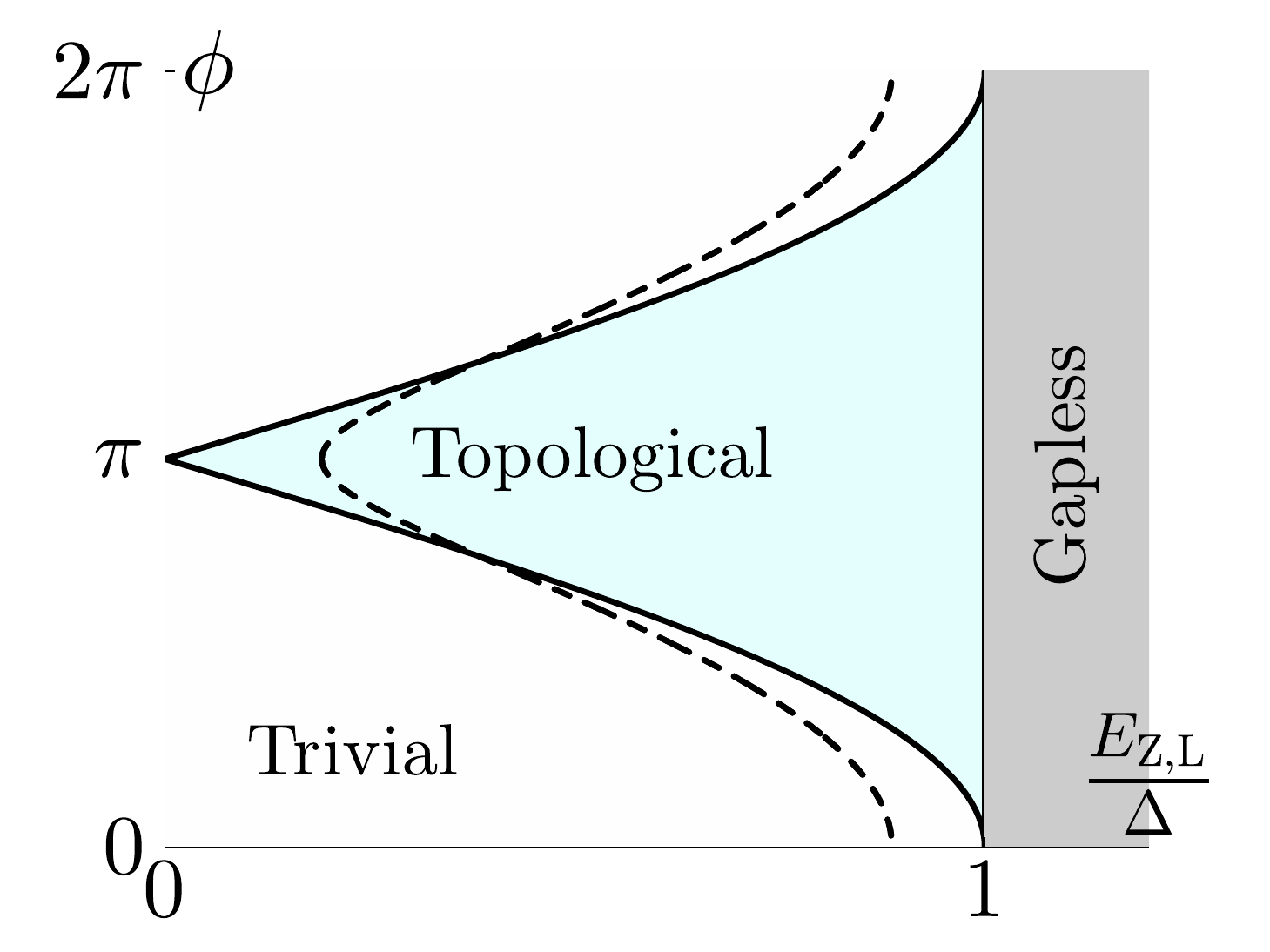}
\caption{Phase diagram as a function of the Zeeman field in the leads $E_{{\rm Z,L}}$
and the phase difference $\phi$ in the limit of a narrow junction
with $E_{{\rm Z,J}}\ll E_{{\rm T}}$. For $E_{{\rm Z,L}}>\Delta$ the system becomes
gapless. The solid lines are the phase boundaries in the absence of
any normal backscattering, while the dashed lines correspond to junction transparency of 0.75 and a phase $k_{{\rm F}}W+\varphi_{N}=3\pi/8$.}
\label{fig:phase_diagram_leads} 
\end{figure}

We have so far neglected the effect of the Zeeman field in the lead, which limits the realization of a topological phase to junctions wide enough that $E_{\rm T}\sim E_{\rm Z,J}$. In the presence of a sizable Zeeman effect in the lead a topological phase may be accessible even in much narrower junctions where $E_{\rm T}$ greatly exceeds experimentally realizable Zeeman fields.
The zero-energy solutions of Eq.~\eqref{eq:bound_states_cond} then read
\begin{equation}
E_{\rm Z,L}=\Delta\cos\left(\frac{\pi}{2}\frac{E_{{\rm Z,J}}}{E_{{\rm T}}}\pm\frac{\phi}{2}\right).\label{eq:phase_boundary_uniform_B}
\end{equation}
In the limit of a narrow junction with $E_{\rm Z,J}\ll E_{\rm T}$, the first term inside the cosine can be neglected. The corresponding phase diagram is shown in Fig. \ref{fig:phase_diagram_leads}. The topological phase is limited to $E_{\rm Z,L}<\Delta$ as larger Zeeman fields drive the leads into a gapless regime.

Inside the gapped regime, the system remains always trivial (topological) at $\phi=0$ ($\phi=\pi$) due to the degeneracy of the spectrum at these values of the phase difference. In the presence of normal reflection the phase difference $\phi$ in Eq.~\eqref{eq:phase_boundary_uniform_B} is replaced by $\tilde{\phi}$ defined in Eq.~\eqref{eq:phi_tilde}. The corresponding phase boundaries are plotted in Fig.~\ref{fig:phase_diagram_leads} as dashed lines.

\subsection{Class BDI Phase Diagram}\label{sec:BDI}

As mentioned in Sec.~\ref{sec:Summary}, the model in Eq.~\eqref{hamil} possesses additional symmetries placing it in the BDI class.
In the absence of a Zeeman field and for a phase difference of $\pi$ between the superconductors, the Hamiltonian is time-reversal symmetric. In addition it commutes with a modified mirror operator with respect to the $x-z$ plane that we define as $\tilde{M}_y=\left(y\to-y\right)\times i\sigma_y\tau_z$. 
A Zeeman field along  $x$, as well as a shift of the phase difference away from $\pi$, breaks both of these symmetries, but remains symmetric to their product. We can therefore define an anti-unitary effective time-reversal operator $\tilde{T}=\tilde{M}_yT$, where $T=i\sigma_yK$ is the standard time-reversal operator with $K$ denoting complex conjugation, which commutes with the Hamiltonian. Note that $\tilde{T}^2=1$. The particle-hole operator, given in the basis we are using by $P=\sigma_y\tau_yK$ obeys $P^2=1$, and we therefore conclude that our model belongs to the BDI symmetry class with a $\mathbb{Z}$ topological invariant. Note that the $\mathbb{Z}_2$ invariant of class D discussed previously is determined by the parity of the $\mathbb{Z}$ invariant. We therefore expect that the topological (trivial) regions found previously will split into subregions with odd (even) $\mathbb{Z}$ indices.

\begin{figure}
\centering 
\includegraphics[width=0.45\textwidth]{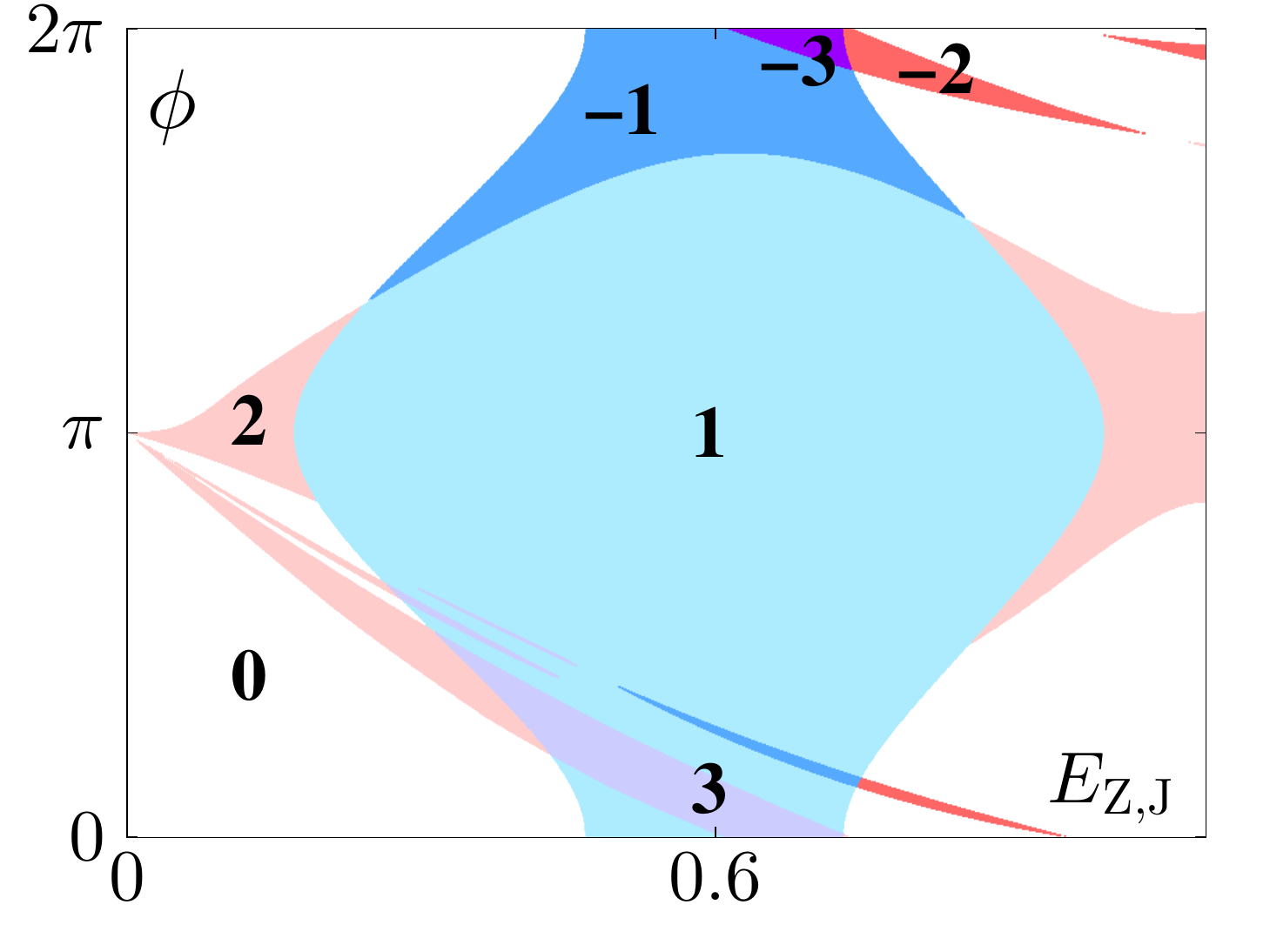}
\caption{A phase diagram of the system as function of the Zeeman field and $\phi$ in presence of the mirror symmetry $\tilde{M}_y$ defined in the text. The values of the $\mathbb{Z}$ invariant corresponding to each region are indicated on the figure. Regions with odd (even) $\mathbb{Z}$ indices, corresponding to topological (trivial) regions of the $D$ class, are filled with shades of blue (red). Note that, by our definition of $\phi$, the phase diagram is not invariant under $\phi\to\phi+2\pi$, since this operation flips the sign of the superconducting gap function. The figure was obtained using a tight binding version of the model (see Appendix \ref{app:Tight-binding}) with the following parameters $W=5$,
$W_{{\rm SC}}=5$, $t=1$, $\alpha=0.5$, $\mu=-2.75$, $\Delta=0.3$. Note that normal reflection is implicitly present in this model due to the finite width of the superconductors.}
\label{fig:BDI_PD}
\end{figure}

To demonstrate this we use a tight binding version of the Hamiltonian in Eq.~\eqref{hamil} (see Appendix \ref{app:Tight-binding} for details of the model) and calculate the BDI invariant following Ref.~\cite{Tewari2012}. To this end we bring the chiral symmetry operator $\tilde{C}=\tilde{M}_y\tau_y$ to a diagonal form with $\mathbb{1}$ in the upper left block and $\mathbb{-1}$ in the lower right block. In this basis the Hamiltonian is purely off-diagonal and we can calculate the phase of the determinant of the off-diagonal part. The invariant is then calculated from the winding of this phase as $k_x$ changes from $0$ to $\pi$. The phase diagram obtained for a particular set of parameters is shown in Fig.~\ref{fig:BDI_PD}. We note that although many additional subregions with various $\mathbb{Z}$ indices appear in the phase diagram, a large $\mathbb{Z}=1$ gapped region is still present.

To stabilize the topological phase with a single Majorana bound state at each end, it is favorable to break this additional symmetry. To this end we introduce different magnitudes for the two superconductors, $\Delta_{1,2}$. When $\left|\Delta_1\right|\neq\left|\Delta_2\right|$, $\tilde{T}$ no longer commutes with the Hamiltonian and the symmetry class is reduced to D. We therefore expect that gap closing lines observed in Fig.~\ref{fig:BDI_PD} corresponding to phase transitions between different $\mathbb{Z}$ invariants with the same parity will no longer be present.

\begin{figure}
\centering 
\includegraphics[width=0.45\textwidth]{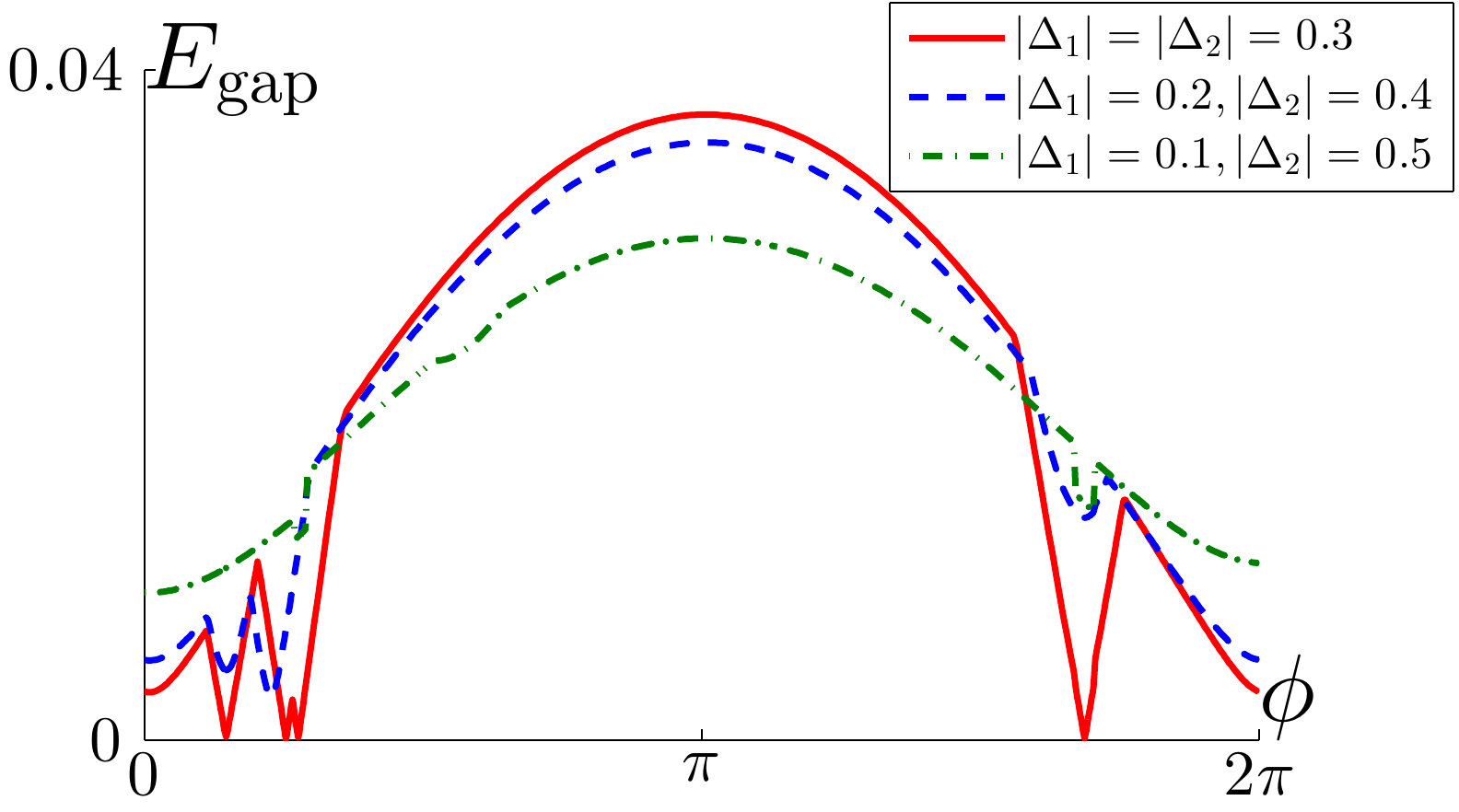}
\caption{Bulk gap calculated along a cut in Fig.~\ref{fig:BDI_PD} with $E_{{\rm Z,J}}=0.6$. When $\left|\Delta_1\right|=\left|\Delta_2\right|=0.3$ the system is in the BDI symmetry class. The bulk gap closes when the system undergoes topological phase transitions between regions with different (odd) $\mathbb{Z}$ indices. When $\left|\Delta_1\right|\neq\left|\Delta_2\right|$ the effective time-reversal symmetry is broken and the bulk becomes gapped for all values of $\phi$.}
\label{fig:BDI_gap}
\end{figure}

To verify this we employ the tight binding model and plot in Fig.~\ref{fig:BDI_gap} the bulk gap versus the phase difference $\phi$ at a constant Zeeman field $E_{{\rm Z,J}}=0.6$, for the same model parameters as in Fig.~\ref{fig:BDI_PD}.
It can be seen that when $\left|\Delta_1\right|=\left|\Delta_2\right|$, the bulk gap closes at values of $\phi$ corresponding to topological phase transitions between different (odd) $\mathbb{Z}$ indices.
Once a different magnitude for the two superconductors is introduced, and the effective time-reversal symmetry is broken, a gap opens for all values of $\phi$.

\subsection{Topological Gap}\label{sec:gap}

\begin{figure*}
\begin{tabular}{lr}
\includegraphics[width=0.72\textwidth]{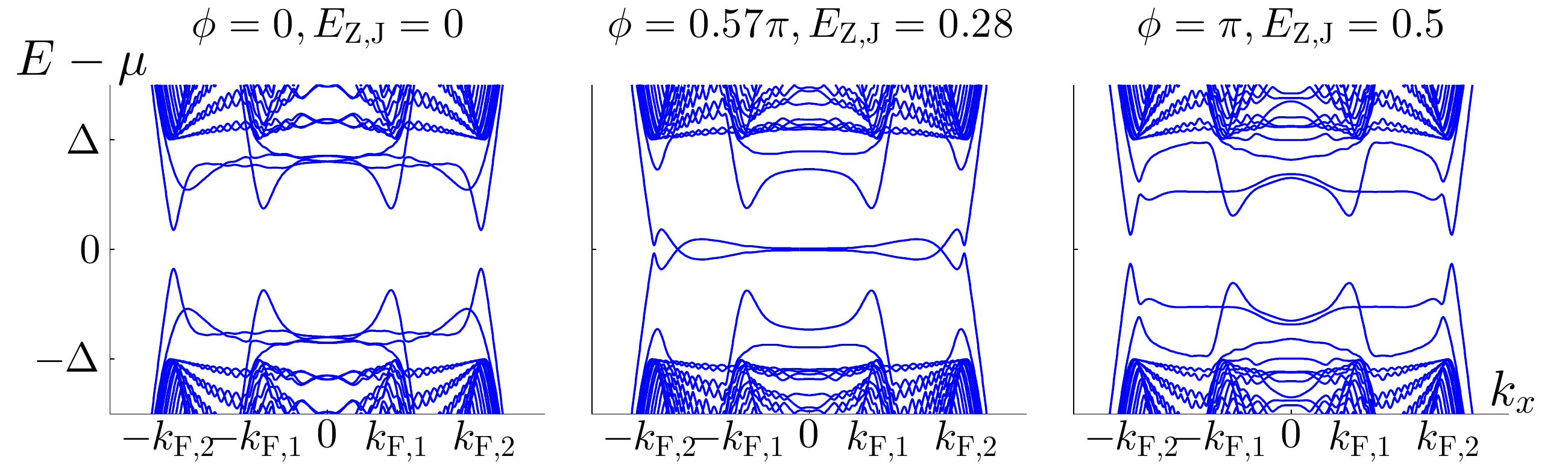}
\llap{\parbox[c]{0.0cm}{\vspace{-7.7cm}
\hspace{-16.5cm}\footnotesize{(a)}
\hspace{3.5cm}\footnotesize{(b)}
\hspace{3.5cm}\footnotesize{(c)}
}} & 
\includegraphics[width=0.28\textwidth]{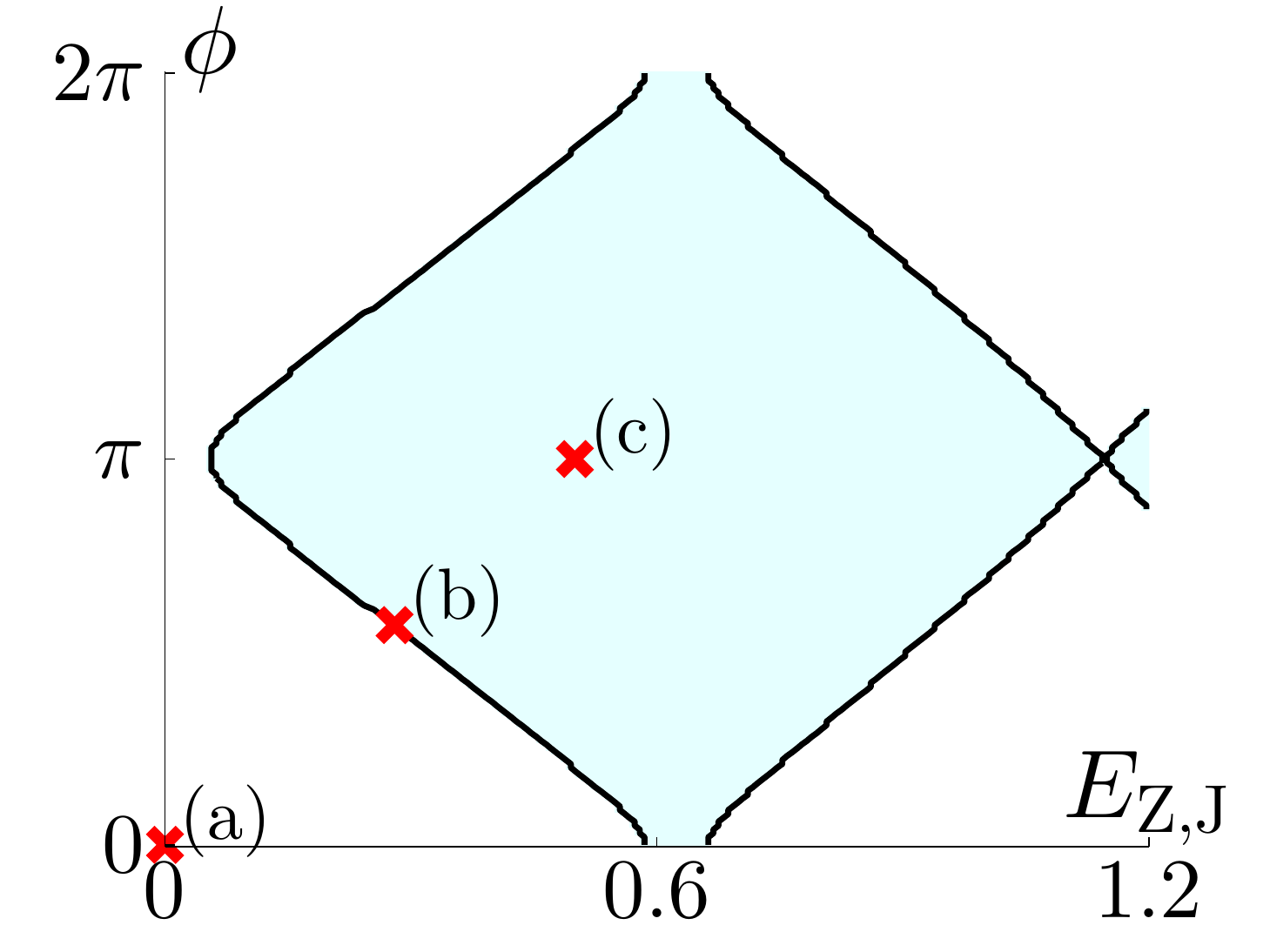}
\llap{\parbox[c]{0.0cm}{\vspace{-7.7cm}\hspace{-10.5cm}\footnotesize{(d)}}}
\end{tabular}

\caption{(a-c) Energy spectrum across the topological phase transition calculated using a tight binding model for the system (see Appendix \ref{app:Tight-binding} for details). The tight binding parameters used are $W=5$, $W_{{\rm SC}}=20$, $t=1$, $\alpha=0.5$, $\mu=-2.8$, $\Delta=0.3$. The Fermi momenta $k_{{\rm F},1/2}$ are
calculated in the absence of a Zeeman field. 
The values of $\phi$ and $E_{{\rm Z,J}}$ for which the spectra are plotted are indicated by crosses on the phase diagram shown in (d). The phase diagram is obtained by calculating the topological invariant for class D, $Q={\rm sign}\left[{\rm Pf}\left(H_{k=\pi}\tau_x\right) / {\rm Pf}\left(H_{k=0}\tau_x\right)\right]$ \cite{Tewari2012}. In (a) the system is in the trivial phsae, in (b) the gap at $k_x=0$ closes and the system undergoes a topological phase transition and in (c) the system is in the topological phase.}
\label{fig:spectrum_vs_kx}
\end{figure*}

\begin{figure}[h]
\includegraphics[width=0.45\textwidth]{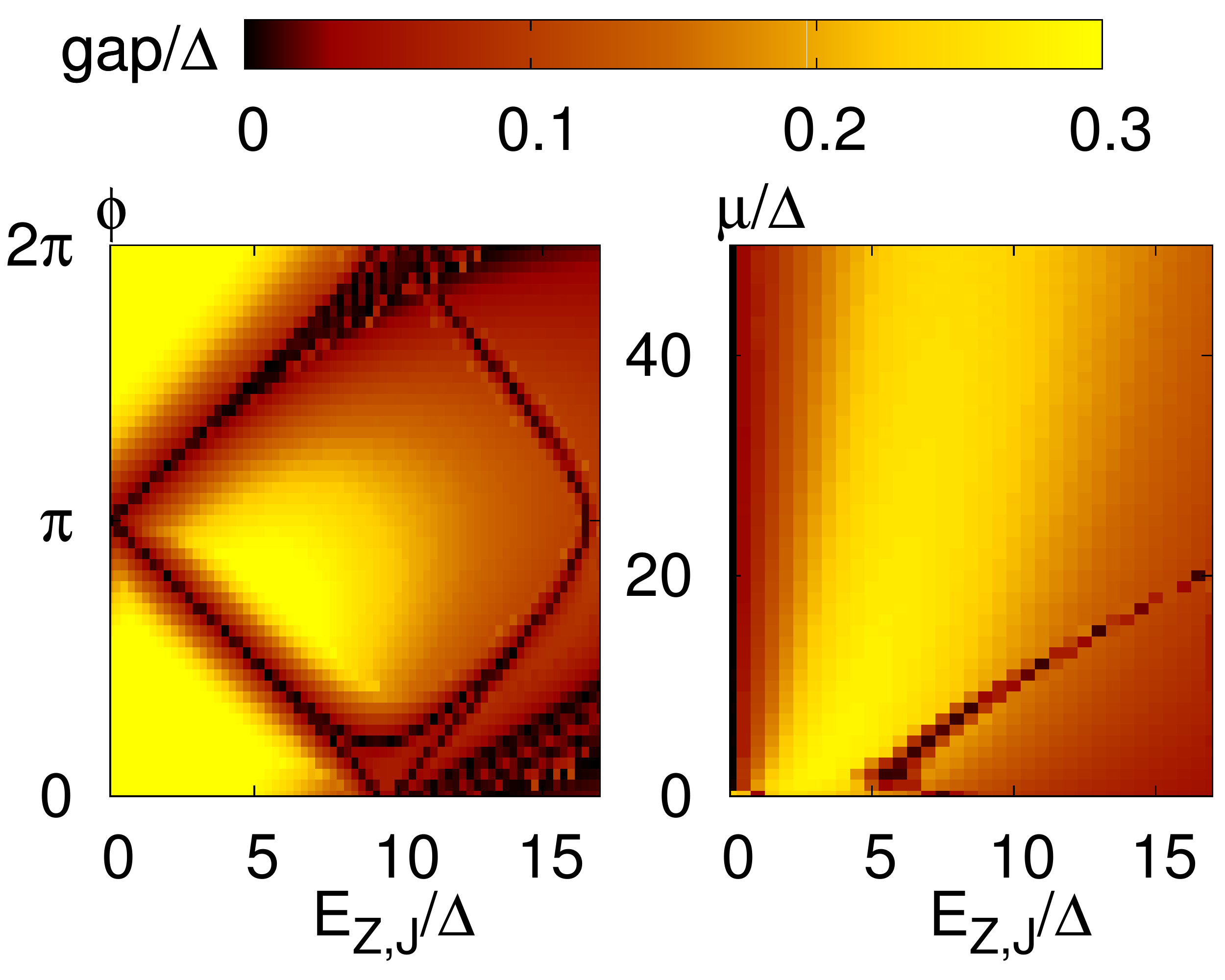}
\caption{Induced gap as function of system parameters evaluated in the continuum model using the scattering matrix approach (see Appx.~\ref{app:ScatteringMatrix} for details) for $W=1(m\Delta)^{-1/2}$, $m\alpha^{2}=9\Delta$, and $E_{\rm Z,L}=0$. In the left panel $\mu/\Delta=20$. The diamond-shaped gap closing lines indicate the boundary between the trivial and the topological regions. Additional regions of small gap occur in the vicinity of BDI phase transitions, where the gap closes at nonzero momenta. In the right panel $\phi=\pi$ and a sizable topological gap is obtained in a very broad range of Zeeman fields with hardly any dependence on the chemical potential.}
\label{fig:gap}
\end{figure}

The topological protection of the phase is governed by the size of
the topological gap, which is determined by the lowest energy Andreev
bound state in the junction. To estimate the magnitude of the gap,
we need to consider the bound state spectrum for all $k_{x}$. Once
again, we consider the case of zero normal
reflection probability and no Zeeman field in the leads $E_{{\rm Z,L}}=0$.

For $k_{x}=0$, the solutions of Eq. \eqref{eq:bound_states_cond} take
a simple form in the two limiting cases of a narrow and wide junction:

\begin{equation}
E=\begin{cases}
\Delta\cos\left(\frac{\pi}{2}\frac{E_{{\rm Z,J}}}{E_{{\rm T}}}\pm\frac{\phi}{2}\right) & \Delta\ll E_{{\rm T}}\\
E_{\rm T} \left(\frac{E_{\rm Z,J}}{E_{\rm T}} \pm\frac{\phi}{\pi} + 2n+1\right) & \Delta\gg E_{{\rm T}}
\end{cases}\label{eq:bs_spectrum}
\end{equation}
The largest gap in the topological region is obtained for $\phi=\pi$
and $E_{{\rm Z,J}}=E_{{\rm T}}$. For a narrow junction the gap is given
by $\Delta$, while for a wide junction the gap is smaller and given
by $E_{\rm T}$. We will consider the scenario of a narrow but
finite width junction with $\Delta\lesssim E_{{\rm T}}$, which is likely to be the most relevant experimentally. Note that in order to
reach the maximal gap in the topological region in this case, a relatively
large Zeeman field $E_{{\rm Z,J}}>\Delta$ is required. 

We next discuss the spectrum for non-zero $k_{x}$. \footnote{Note that the
spectrum is symmetric in $k_{x}$ due to mirror-symmetry with respect to the $y-z$ plane, and hence we can consider only $k_{x}>0$.}
In this case, spin is no longer a good quantum number, and spin-orbit coupling can
not be gauged out. We denote the magnitude of the Fermi momentum on
the inner (outer) Fermi surface in presence of spin-orbit coupling
by $k_{{\rm F},1(2)}=k_{\rm F}\mp k_{\rm SO}$, where $k_{\rm F}=\left(2 m \mu\right)^{1/2}$ and $k_{\rm SO}=m\alpha$. For a given $k_{x}$, we denote the $y$ component
of the Fermi momenta on the two Fermi surfaces by $k_{{\rm F},i,y}=\left(k_{{\rm F},i}^{2}-k_{x}^{2}\right)^{1/2}=k_{{\rm F},i}{\rm sin}\theta_{i}$,
where $\theta_{i}={\rm cos}^{-1}\left(k_{x}/k_{{\rm F},i}\right)$. We
note that for a given $k_{x}$, the spins of the electrons on the
two Fermi surfaces are no longer orthogonal. Therefore, when, e.g.,
an electron in the vicinity of the inner Fermi surface is Andreev reflected
from the superconductor, it will be reflected as a superposition of
holes from both the inner and the outer Fermi surfaces. 
However, in the limit of small spin-orbit coupling, $\alpha k_{\rm F}\ll\mu$, the overlap between the spins on the different Fermi surfaces remains small. (This is assuming that the Zeeman field does not alter the Rashba induced spin-momentum locking, i.e. $E_{\rm Z,J}\ll\alpha k_{\rm F}$.) In the opposite limit of large spin-orbit coupling, a large momentum transfer $\delta k_y = k_{{\rm F},2,y}-k_{{\rm F},1,y}$ is required for such a process. If $\delta k_y \gg \Delta / v_{{\rm F},1,y}$, such scattering is suppressed.
We conclude that Andreev reflection between different Fermi surfaces can be neglected if $E_{\rm Z,J},\Delta\ll\alpha k_{\rm F}$. In this case the scattering equations for the two spin species (corresponding to the two Fermi surfaces) can still be decoupled.

To write down the scattering equation we need to determine the phase shift acquired by an electron (or a hole) upon crossing the normal region of the 2DEG. 
To this end, we use the plane wave basis along $y$ and diagonalize the Hamiltonian \eqref{hamil} in the normal region $\left|y\right|<W/2$. The resulting spectrum for the electrons is given by
\begin{equation}
E=\frac{k_{x}^{2}}{2m}+\frac{k_{y}^{2}}{2m}-\mu+\frac{m\alpha^{2}}{2}\pm\sqrt{\alpha^{2}k_{x}^{2}+\left(E_{{\rm Z,J}}-\alpha k_{y}\right)^{2}}.\label{eq:normal_spectrum_B}
\end{equation}
The spectrum for the holes can be obtained using particle-hole symmetry.
We see that the energy shift of an electron (or a hole) on Fermi surface $i$ due to the Zeeman field, to first order in $E_{{\rm Z,J}}$, is given by $\Delta E_{i}\simeq E_{{\rm Z,J}}{\rm sin}\theta_{i}$.
Therefore, the phase accumulated
when traversing the junction, $\left(\Delta E_{i}/v_{{\rm F},i,y}\right)W=\left(E_{{\rm Z,J}}/v_{{\rm F}}\right)W$,
is the same for the two Fermi surfaces and is independent of $k_{x}$.

We conclude that the scattering equation for the bound states at nonzero $k_x$ is given by Eq.~\eqref{eq:bound_states_cond} with $E_{{\rm T}}\to E_{{\rm T},i}\left(k_{x}\right)=\left(\pi/2\right)\left(v_{{\rm F},i,y}/W\right)$ and $E_{{\rm Z,J}}\to E_{{\rm Z,J}}\sin\theta_i$ (such that the ratio $E_{{\rm Z,J}}/E_{{\rm T}}$ is left unchanged). Hence, the energies are given by Eq.~\eqref{eq:bs_spectrum} with the same substitution.
Note that the $k_{x}$-dependent Thouless energy decreases with increasing
$k_{x}$. Once $E_{{\rm T},i}\left(k_{x}\right)$ becomes smaller
than $\Delta$, multiple bound states appear and the gap at $k_x$ becomes governed by $E_{{\rm T},i}\left(k_{x}\right)$.
As $k_{x}$ approaches $k_{{\rm F},i}$ the gap is reduced to be of
order $1/\left(mW^{2}\right)$. 

At high values of the Zeeman field, the gap can be further limited by the following effect. In the normal state, $\Delta=0$, and in the presence of a non-uniform Zeeman field, $E_{\rm Z,J} > E_{\rm Z,L}$, 
a potential well is formed by the Zeeman energy in the normal region. The depth of this potential depends on $k_x$. In the limit $E_{\rm Z,J}\ll\alpha k_{\rm F}$, the potential at $k_x\simeq k_{\rm F}$ is equal to $-E_{{\rm Z,J}}^{2}/\left(\alpha k_{{\rm F}}\right)$, as can be seen from Eq.~\eqref{eq:normal_spectrum_B}.
States at momenta close to $k_{\rm F}$, bound by this potential, can lead to a suppression of the superconducting gap, once the characteristic length for the decay of their transverse wavefunction, $\xi_B$, becomes smaller than $W$. In this regime, the decay length is given by $\xi_{B}\simeq\left[2m E_{{\rm Z,J}}^{2}/\left(\alpha k_{{\rm F}}\right)\right]^{-1/2}$. 
The discussion above suggests, that the optimal gap in the system is obtained at $\phi=\pi$ and $E_{\rm Z,J}=E_{\rm T}$ and is equal to ${\rm min}\left\{ \Delta, \left(1/mW^2\right) \right\}$. However, if the Zeeman field for which $\xi_B$ becomes smaller than $W$ is smaller than $E_{\rm T}$, the optimal gap can be suppressed.
The value of the Zeeman field at which $\xi_B\sim W$ is given by $E_{\rm Z,J}^{\rm c}=E_{\rm T}\left(\alpha / v_{\rm F}\right)^{1/2}$.
This allows us to obtain a lower bound on the optimal gap in the system. In a narrow junction, $\Delta\lesssim 1/\left(mW^2\right)$, the gap at $\phi=\pi$ for $E_{\rm Z,J}<E_{\rm Z,J}^{\rm c}$ is given by $\Delta\sin\left[\left(\pi/2\right)E_{\rm Z,J}/E_{\rm T}\right]$. At $E_{\rm Z,J}=E_{\rm Z,J}^{\rm c}$ this gives a gap of order $\Delta\left(\alpha / v_{\rm F}\right)^{1/2}$.
Thus a gap of order $\Delta$ can be reached for $v_{\rm F}\lesssim \alpha$ but due to the slow, power-law dependence, the system has a sizable gap also for larger values of $v_{\rm F}$.
In a wide junction, $\Delta\gg 1/\left(mW^2\right)$, the superconducting gap will, in fact, be effected by a finite $\xi_B$ only once it becomes smaller than $\xi=1/\left(m W\Delta\right)\ll W$. This leads to a much looser constraint on the Fermi velocity, allowing for a gap of order $1/\left(mW^2\right)$ as long as $v_{\rm F}/\alpha<W/\xi$.

We conclude that a topological gap of order $\Delta$ can be obtained
if the junction is narrow, $\Delta\lesssim 1/\left(mW^{2}\right)$, and the
chemical potential is such that $v_{{\rm F}}\lesssim\alpha$.
For a wider junction the size of the topological gap is governed by
$1/\left(mW^{2}\right)$. The optimal gap is obtained for $\phi=\pi$ and $E_{{\rm Z,J}}\lesssim E_{{\rm T}}$.

To complement this analysis we calculate the spectrum of the system as function of $k_x$ across the phase transition, using a tight binding version of the model, given in Appendix \ref{app:Tight-binding}, and plot it in Fig.~\ref{fig:spectrum_vs_kx}. It can be seen that both in the trivial and the topological regions the smallest gap occurs at $k_x \simeq k_{{\rm F},i}$, in agreement with the discussion above. At the phase transition, the gap at $k_x=0$ closes, and it is in fact close to zero also for other $k_x$.

In addition we calculate the gap numerically using
the scattering matrix approach (see Fig. \ref{fig:gap}). We consider
a narrow junction with $\Delta\simeq 1/\left(mW^{2}\right)$ and find that
a sizable gap of order $\Delta$ can indeed be obtained for $\phi=\pi$ with very weak dependence on the chemical potential.

\subsection{Majorana End Modes}

\begin{figure}
\includegraphics[width=0.49\textwidth]{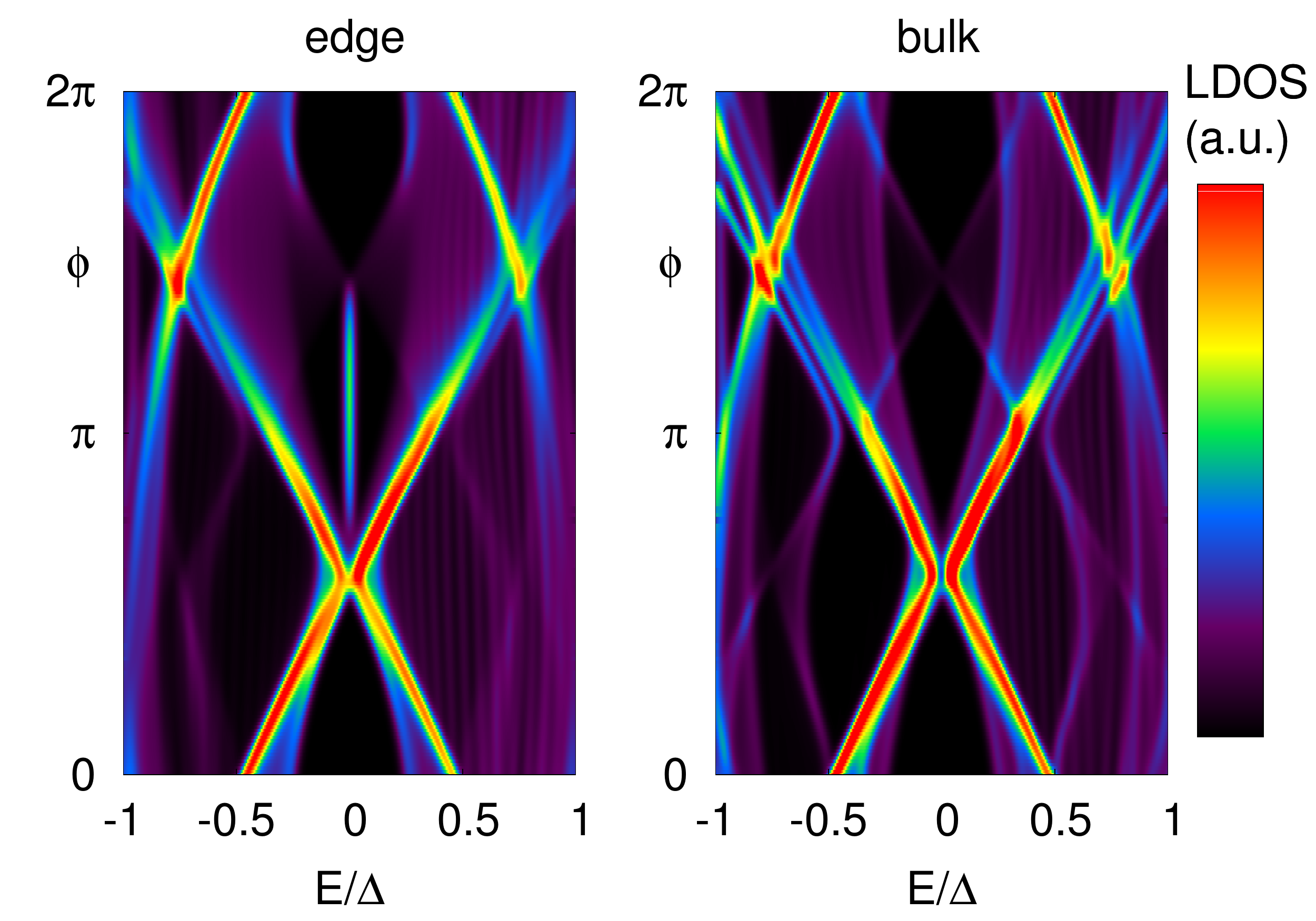}

\caption{Local density of states at the edge (left panel) and in the center (right panel) of the junction as a function
of energy and phase difference. In a range around $\phi=\pi$ a Majorana
state forms at the edge. The result is obtained numerically from a tight-binding
model (see Appendix \ref{app:Tight-binding}) using the following
parameters (energies and length are in units of the hopping strength
and lattice spacing): $\alpha=0.5$, $E_{{\rm Z,J}}=E_{{\rm Z,L}}=0.1$,
$\Delta=0.25$, $\mu=-3.75$ (measured from the center of the tight-binding
band), junction width $W=4$, width of the superconducting leads $W_{{\rm SC}}=8$,
length $L=200$. We plot a spatial average of the density of states
over a rectangle spanning the entire width of the junction in the
$y$ directions and the first 10 sites from the edge (left panel) or the most central 10 sites (right panel) in the $x$ direction.
For presentation the local density of states has been convoluted with
a Gaussian with a standard deviation of $0.02\Delta$. \label{fig:ldos}}
\end{figure}

In the topological phase we expect the system to host Majorana bound states at its ends. To verify the appearance of these zero-energy bound states in the proposed setup we calculate the local density of states (LDOS) close to the boundaries of the system. To this end we diagonalize a tight-binding version of the Hamiltonian in Eq.~\eqref{hamil} with boundaries both along the $x$ and the $y$ dimensions (for details of the model see Appendix \ref{app:Tight-binding}). The resulting LDOS as function of the phase difference is shown in Fig. \ref{fig:ldos}. Indeed, a zero-energy state is present at the end of the junction in a finite range of phase differences around $\phi=\pi$. 

Note that in presence of the effective time-reversal symmetry discussed in Sec.~\ref{sec:BDI}, multiple Majorana bound states will appear at each end of the system. The number of zero-energy states in this case will be determined by the BDI $\mathbb{Z}$ invariant.

\section{First order topological phase transitions and the critical current}\label{sec:critical_current}

\begin{figure*}
\centering
\includegraphics[clip=true,trim =1cm 0cm 0cm 0cm,width=0.9\textwidth]{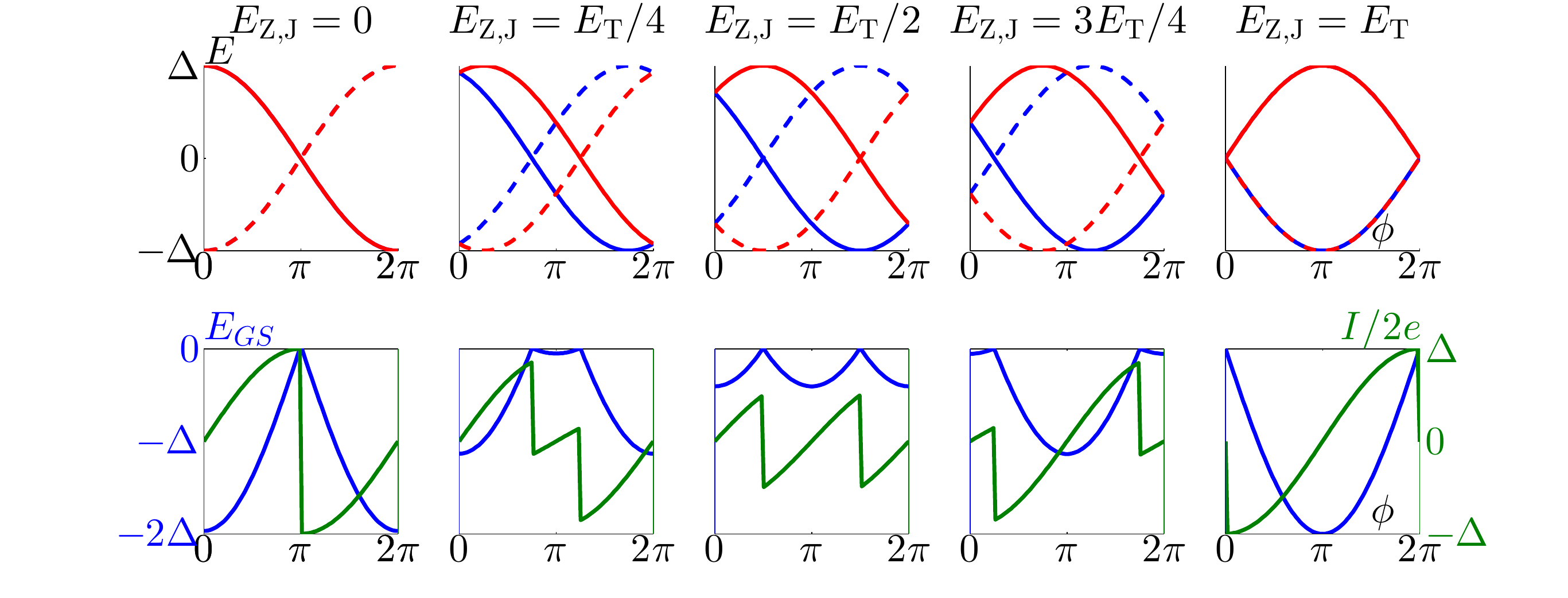}
\caption{The upper panel shows the bound state energies of the two spin species (plotted in red and blue) and the energies of their particle-hole symmetric states (indicated by dashed lines) for a single momentum $k_x<k_{{\rm F},1}$ in a narrow junction, as the Zeeman field is varied. The contribution to the ground state energy (obtained by summing over the negative energy states) and the Josephson current are plotted for each value of Zeeman field in the lower panel in blue and green, respectively. At $E_{{\rm Z,J}}=E_{{\rm T}}/2$
the value of $\phi$ for which the energy is minimized shifts from $0$ to $\pi$. This transition
is accompanied by a minumum in the critical current. \label{fig:BoundStates_and_Egs_Narrow}}
\end{figure*}

In this section we show that if the phase difference is not imposed externally, the system will self-tune into the topological phase in a wide range of Zeeman fields. 
Using the bound state spectrum obtained in Sec.~\ref{sec:Phase-Diagram-and-Gap} we calculate the ground state energy of the system and the Josephson current in the junction. At a critical value of the Zeeman field the system undergoes a first order phase transition, in which the ground state of the junction switches between values of $\phi$ corresponding to the trivial and the topological phases, and that this transition is accompanied by a minimum of the critical current.

To this end, we need to sum over the contributions of all $k_{x}$ to the ground state energy. 
In the analysis of the gap presented in Sec.~\ref{sec:gap} we found that in the limit $\Delta\ll\mu$, as well as $E_{{\rm Z,J}}\ll\alpha k_{{\rm F},1/2}$, and assuming $E_{\rm Z,L}=0$ and no normal reflection, the spectrum for $k_{x}<k_{{\rm F},1}$ is given by Eq.~\eqref{eq:bs_spectrum}, with $E_{{\rm T}}\to E_{{\rm T},i}\left(k_{x}\right)$ and $E_{{\rm Z,J}}\to E_{{\rm Z,J}}\sin\theta_i$ (such that the ratio $E_{{\rm Z,J}}/E_{{\rm T}}$ is left unchanged).
For $k_{{\rm F},1}<k_{x}<k_{{\rm F},2}$, there is only a single spin species present in the system and thus only half of the bound states remain.

We first calculate the ground state energy and the critical current in the limit $\alpha k_{{\rm F}}\ll\mu$.
In this limit $\left( k_{{\rm F},2}-k_{{\rm F},1} \right) / k_{{\rm F}} = 2 k_{\rm SO} / k_{\rm F} \to 0$, and we can therefore neglect the contribution of momenta in the range $k_{{\rm F},1}<k_{x}<k_{{\rm F},2}$. We later relax this constraint and discuss how the results are altered.

We first focus on the limit of an ultra-narrow junction with a single bound state (for each spin species) for all $k_x$, i.e. $\Delta\ll 1/\left(mW^2\right)$, and consider the contribution of a single $k_x$ to the ground state energy. In this limit, the dominant contribution to the $\phi$-dependent part of the ground state energy (and thus also to the Josephson current) comes from the Andreev bound states \cite{Beenakker1991}. We denote the bound state energies by $E_{\pm}=\Delta\cos\left(\phi_{B} \pm \phi/2\right)$ , where $\phi_{B}=\left(\pi/2\right)E_{{\rm Z,J}}/E_{{\rm T}}$ is used as a shorthand notation (note that $\phi_{B}$ is independent of $k_x$). The ground state energy is obtained by summing over the negative energy states, i.e. $E_{{\rm GS}}=-\left|E_{+}\right|-\left|E_{-}\right|$. The spectra of the bound states as well as the resulting ground state energy are plotted in Fig. \ref{fig:BoundStates_and_Egs_Narrow} for different values of $\phi_{B}$.
We note that at $E_{{\rm Z,J}}=\left(n+1/2\right) E_{{\rm T}}$ the value of $\phi$ for which the energy is minimized switches between $\phi=0$ and $\phi=\pi$.
Since the energy dependence on $\phi$ and the Zeeman field in this case is the same for all $k_x$, we conclude that at $E_{{\rm Z,J}}=\left(n+1/2\right) E_{{\rm T}}$ the ground state of the entire system switches from between $\phi=0$ and $\phi=\pi$. Note that in this transition the fermion parity of the $k_x=0$ mode changes, indicating a transition into the topological phase. This is a first order phase transition without a gap closing.

We next calculate the critical current in the junction in the same limit. At zero temperature, the Josephson current is given by $I\left(\phi\right)=2e \frac{d}{d\phi} E_{{\rm GS}}$.
We note that the maximum of the Josephson current is obtained at the same value of $\phi$ for all $k_x$. We can therefore calculate the critical current in the system, $I_{c}={\rm max}\left|I\left(\phi\right)\right|$, as function of the Zeeman field, based on a single $k_x$.
Due to the relative phase shift in the bound-states spectra of the
two spins, the critical current of a single momentum is modulated as the Zeeman field is varied and is equal to $I_{c}=2e\Delta{\rm max}\left\{ {\rm cos}^{2}\phi_{B},{\rm sin}^{2}\phi_{B}\right\} $ (see lower panel of Fig.~\ref{fig:BoundStates_and_Egs_Narrow}).
The maximal value, $I_{c,{\rm max}}=2e\Delta$ is obtained for $\phi_{B}=\pi n/2$, or equivalently $E_{{\rm Z,J}}=n E_{{\rm T}}$
and the minimal one, $I_{c,{\rm min}}=I_{c,{\rm max}}/2=e\Delta$
is obtained for $\phi_{B}=\left(\pi/2\right)\left(n+1/2\right)$, or equivalently $E_{{\rm Z,J}}=\left(n+1/2\right) E_{{\rm T}}$
\cite{Nazarov2014}. 
Note that the minima of the critical current occur exactly 
at $0-\pi$ transitions of the junction. 
The value of the Zeeman field at which this transition takes place is in agreement with the semiclassical argument given in Sec. \ref{sec:Summary}. However, we see that the critical current does not vanish at these points.

When a finite temperature is considered, the Josephson current is
given by $I\left(\phi\right)=2e \frac{d}{d\phi} F$, where $F$ is the
free energy of the system. In the high temperature limit, $T\gg\Delta$,
we obtain (see Appendix \ref{app:Finite-temperature})
\begin{equation}
I\left(\phi\right)\simeq-\frac{4e}{T}\sum_{n}E_{n}\frac{dE_{n}}{d\phi}=2e\frac{\Delta^{2}}{T}\cos\left(2\phi_{B}\right)\sin\left(\phi\right).\label{eq:NarrowHighT}
\end{equation}
We find that only the first harmonic of the Josephson current is left.
The critical current is proportional to ${\rm cos}\left(2\phi_{B}\right)$ and is hence zero for $\phi_{B}=\left(\pi/2\right)\left(n+1/2\right)$. The suppression of the minimum of the critical current with temperature can be seen in Fig.~\hyperref[fig:Ic]{\ref{fig:Ic}(b)}.
The semiclassical result is thus recovered in the high temperature limit. This is due to the fact that higher harmonics of the critical current, which correspond to multiple Andreev reflections in the junction that are not accounted for in the semiclassical argument, are suppressed in the high temperature limit.

We next lift the constraint of an ultra-narrow junction. More specifically, we assume that for some $k_x$, $\Delta\ll E_{{\rm T},i}\left(k_{x}\right)$. In the limit of $k_{\rm SO}\ll k_{\rm F}$, we have $E_{{\rm T},1}\left(k_{x}\right)\simeq E_{{\rm T},2}\left(k_{x}\right)$ and we therefore suppress the band index below.
The maximal supercurrent is still obtained at the same value of $\phi$ for all $k_x$ as will be clear from the analysis below. Therefore, we can once again calculate the contribution of a single momentum $k_x$ to the critical current of the system, by considering the supercurrent due to that momentum only.
In this case the contribution of the states above the gap to the energy and the Josephson current can not in general be neglected \cite{ishii1970josephson}.
However, taking the limit $\Delta\to\infty$ allows us to consider
only the bound states. Following the derivation in Ref.~\cite{zagoskin2014book},
we find that the Josephson current in presence of the Zeeman field
is given by
\begin{equation}
I\left(k_x, \phi\right)=8eT\sum_{p=1}^{\infty}\left(-1\right)^{p+1}\frac{\cos\left(2p\phi_{B}\right)\sin\left(p\phi\right)}{\sinh\left(\pi^{2}pT/E_{\rm T}\left(k_x\right)\right)}.\label{eq:CurrentWide-1}
\end{equation}

At zero temperature we obtain
\begin{equation}
I\left(k_x, \phi\right)=\frac{8e}{\pi^{2}}E_{\rm T}\left(k_x\right)\sum_{p=1}^{\infty}\left(-1\right)^{p+1}\frac{{\rm cos}\left(2p\phi_{B}\right){\rm sin}\left(p\phi\right)}{p}.
\end{equation}
We note that since $E_T\left(k_x\right)$ decreases with increasing $k_x$ the contribution of larger $k_x$ to the critical current is smaller. For $\phi_{B}=\pi n/2$ the sum over $p$ converges to a $2\pi$
periodic sawtooth function, and the critical current is maximal and
equal to $I_{\rm c,max}\left(k_x\right)=4e E_{\rm T}\left(k_x\right)/\pi$. For $\phi_{B}=\left(\pi/2\right)\left(n+1/2\right)$,
all the odd harmonics are absent and we obtain a $\pi$ periodic sawtooth
function of half the amplitude, i.e. $I_{\rm c, min}\left(k_x\right)=I_{\rm c, max}\left(k_x\right)/2=2e E_{\rm T}\left(k_x\right)/\pi$.
Note that also in this case the minima of the critical current occur at the values of the Zeeman field for which the minimum of the energy switches between being at $\phi=0$ and $\phi=\pi$, as can be seen by integration of the Josephson current over $\phi$.

In the high temperature limit, which in this case corresponds to $T\gg E_{\rm T}\left(k_x\right)$,
once again only the first harmonic is left:

\begin{equation}
I\left(k_x, \phi\right)=4eTe^{-\pi^2T/E_T\left(k_x\right)}\cos\left(2\phi_{B}\right)\sin\left(\phi\right),\label{eq:WideHighT-1}
\end{equation}
resulting in a vanishing current for $\phi_{B}=\left(\pi/2\right)\left(n+1/2\right)$.
Note also that the critical current contribution from larger $k_x$ is suppressed more strongly at finite temperatures.

\begin{figure}
\centering
\includegraphics[width=0.5\textwidth]{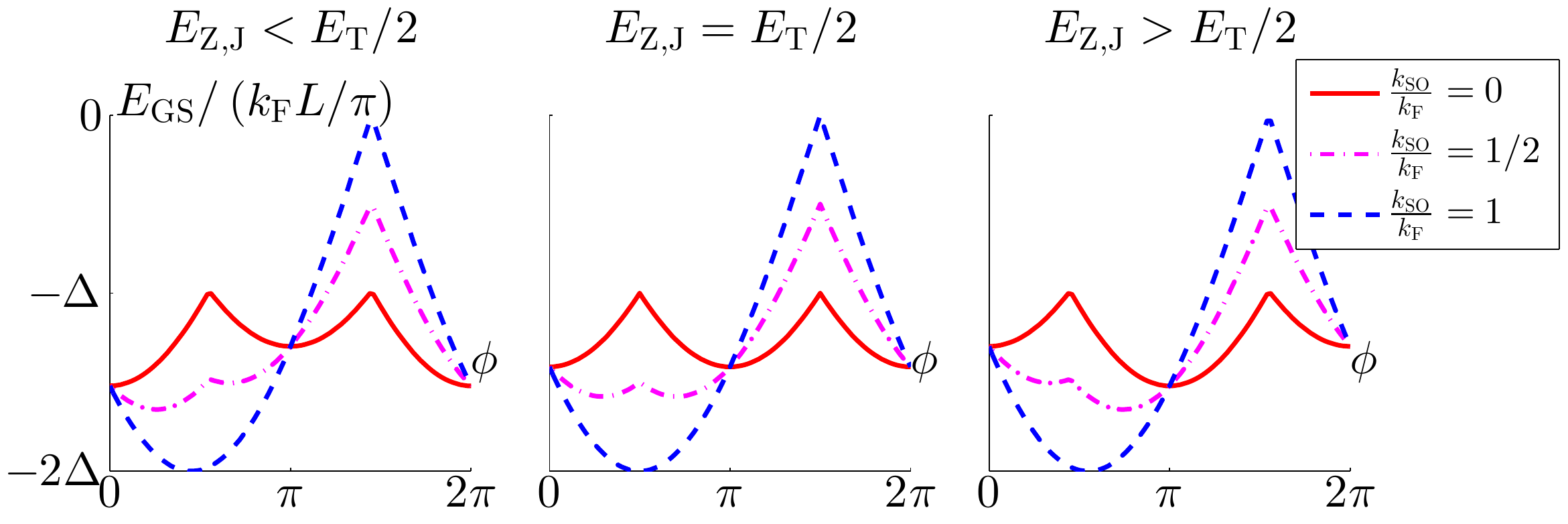}
\caption{The ground state energy of the junction as function of the phase difference $\phi$ in the limit of a narrow junction with a single bound state for all $k_x$ (see Eq.~\eqref{eq:Egs_sumkx}). For $k_{{\rm F}}>k_{{\rm SO}}$, at $E_{{\rm Z,J}}=E_{{\rm T}}/2$, the system undergoes a first order phase transition as the ground state shifts between the two local minima. In the left panel $E_{{\rm Z,J}}=0.45 E_{{\rm T}}$ and the system is in the trivial phase in its ground state. In the right panel $E_{{\rm Z,J}}=0.55 E_{{\rm T}}$ and the system is in the topological phase in its ground state.}
\label{fig:Egs_phi}
\end{figure}

We now lift the constraint $\alpha k_{{\rm F}}\ll\mu$, and consider the contribution of momenta $k_{{\rm F},1}<k_{x}<k_{{\rm F},2}$ to the ground state energy. For simplicity, in this analysis, we will once again consider the limit of an ultra-narrow junction, $\Delta\ll 1/(mW^2)$. For $k_{x}>k_{{\rm F},1}$, there is a single spin species in the system and the energy of the corresponding bound state is given by $E_{-}=\Delta\cos\left(\phi_{B} - \phi/2\right)$ (assuming $\alpha>0$). 
Upon integration of the energy over $k_x$ from $-k_{{\rm F},2}$ to $k_{{\rm F},2}$ we obtain
\begin{equation}
\begin{split}
& E_{{\rm GS}} =   \\ & -\frac{\Delta L}{\pi}\left(\left|\cos\left(\frac{\phi}{2}+\phi_{B}\right)\right|k_{{\rm F},1}+\left|\cos\left(\frac{\phi}{2}-\phi_{B}\right)\right|k_{{\rm F},2}\right).
\end{split}
\label{eq:Egs_sumkx}
\end{equation}
This function is depicted in Fig.~\ref{fig:Egs_phi} for several values of $E_{\rm Z,J}$ and $k_{\rm SO}/k_{\rm F}$.
Concentrating on $\phi_B\leq\pi/2$ and $\phi\leq\pi$, we find that this function can have two local minima at $\phi=\phi_{1,2}$ given by
\begin{equation}
\begin{split}
\tan\frac{\phi_1}{2}&=\tan\phi_{B}\frac{k_{{\rm SO}}}{k_{{\rm F}}}  \hspace{2.3cm} 0\leq\phi_1\leq\pi-2\phi_{B} \\
\cot\frac{\phi_2}{2}&=\frac{k_{{\rm SO}}}{k_{{\rm F}}+\left(\tan\phi_{B}-1\right)k_{{\rm SO}}}  \hspace{0.5cm} \pi-2\phi_{B}\leq\phi_2\leq\pi.
\end{split}
\end{equation}

At $\phi_B=\pi/4$, or equivalently $E_{{\rm Z,J}}=E_{{\rm T}}/2$, it can be shown that $\phi_1+\phi_2=\pi$, and that $E_{{\rm GS}}\left(\phi_1\right)=E_{{\rm GS}}\left(\phi_2\right)$. We can therefore conclude that at this value of the Zeeman field a first order phase transition occurs with the value of $\phi$ changing abruptly between $\phi_1$ and $\phi_2$. 
As long as $k_{{\rm F},1}>0$, or equivalently $k_{{\rm F}}> k_{{\rm SO}}$, we have $\phi_1<\pi/2$ and $\phi_2>\pi/2$ at the transition point. Therefore the system is in the trivial phase on one side of the transition and is in the topological phase on its other side. 

Finite temperature will smoothen the cusp in the ground state energy of the system as function of the phase difference. However, for low enough temperatures two local minima in the free energy still exist allowing for a first order phase transition between them as the Zeeman field is varied.

We note also that in presence of normal reflection the values of the Zeeman field for which the phase transitions will occur, as well as the values for which the critical current will be minimal, will generically shift away from $E_{{\rm Z,J}}=\left(n+1/2\right)E_{{\rm T}}$ and might no longer coincide. However, as long as the normal reflection probability is not too large, we expect these deviations to be small.

Tight-binding calculations of the critical current, complementing this analysis, were presented in Fig.~\hyperref[fig:Ic]{\ref{fig:Ic}(b)}. Note that these calculations are performed in a different regime, where $E_{\rm Z,J} \gtrsim \alpha k_{\rm F}$. Nevertheless, we find that a first order topological phase transition still occurs from $\phi_{\rm GS}$ close to zero to $\phi_{\rm GS}$ close to $\pi$. 

\section{Discussion\label{sec:Discussion}}

We have shown that one-dimensional topological superconductivity can be realized in a Josephson junction across a 2DEG with Rashba spin-orbit coupling and in-plane magnetic field. Once the phase difference between the superconductors is set to $\pi$, a ballistic junction is driven into the topological phase without any further fine tuning. If the phase is not set externally, the system can self tune into the topological phase for a range of in-plane magnetic fields. In this case the modulation of the critical current serves as a diagnostics of the phase transitions.

In practice, the system parameters should be chosen in a way to optimize the gap $\Delta_{\rm top}$ protecting the topological phase. We find that narrow junctions with $1/\left(mW^2\right)\simeq \Delta$ allow for a gap of order $\Delta$ in the topological phase. To reach this limit the chemical potential should ideally not exceed the spin-orbit energy $m\alpha^2$, although we find sizable gaps even for larger values of $\mu$, as the gap decays at most as $\Delta_{\rm top}\sim \Delta (\alpha/v_{{\rm F}})^{1/2}$.
The width of the junction also dictates the magnitude of the Zeeman field required to be close to the center of the topological phase, $E_{{\rm Z,J}}\lesssim E_{{T}}$, i.e. a large Zeeman field is required if the junction is narrow. Moreover, we assume in our estimate of the gap that the Zeeman field does not interfere with the Rashba-induced spin-momentum locking, i.e. that $E_{{\rm Z,J}}\ll\alpha k_{{\rm F}}$. Thus materials with large spin-orbit coupling are favorable.

Although the orbital effects of the in-plane field have not been discussed in the manuscript, we note that nonzero magnetic field in the region between the superconducting leads and the 2DEG can give rise to a spatial modulation of the superconducting order parameter and destroy the gap in the system. Moreover, this effect can lead to oscillations of the critical current as  function of the magnetic field that are not of topological origin. We further elaborate on this in Appendix \ref{app:Orbital}.

While we expect the topological phase to be stable to a certain amount of disorder, the system will eventually enter a trivial phase at strong disorder. It would be interesting to compare the effects of disorder with topological superconductors based on semiconductor nanowires. The latter are restricted to relatively small chemical potentials, where the effects of disorder is particularly severe. This indicates that topological phases in planar Josephson junctions, for which this restriction does not exist, could be more resilient to disorder.

{\it Note added:} While we were preparing this manuscript we became aware of Ref.~\onlinecite{Hell2016} which discusses topological superconductivity in a similar setup, as well as Ref.~\onlinecite{Sticlet2016} which analyzes surface states of nanowires with some relation to our results.

\section*{Acknowledgments}
We acknowledge stimulating discussions with Y.\ Oreg, C.\ Marcus, F.\ Nichele.
E.\ B.\ was supported by the Minerva foundation, by a Marie Curie Career Integration Grant (CIG), and by the European Research Council (ERC) under the European Union's Horizon 2020 research and innovation
programme (grant agreement No.\ 639172). F.\ P., A.\ Y., and B.\ I.\ H.\ acknowledge financial support by the STC Center for Integrated Quantum Materials, NSF Grant No.\ DMR-1231319. A.\ Y. was also supported by the NSF Grant No.\ DMR-1206016. A.\ S. acknowledges financial support by the European Research Council (ERC) under project MUNATOP, Microsoft Station Q, and Minerva foundation.


\appendix

\section{Phase diagram with normal reflection\label{app:normal_reflection}}

We now turn to an estimate of the phase boundaries in the presence of normal reflection. We assume that the mean free path exceeds the width of the junction so that normal reflection is limited to the superconducting regions of the 2DEG and to the superconducting-normal interface. Normal reflection can arise when $\mu$ and $\Delta$ are of the same order or when the width of the superconducting segment is comparable to the superconducting coherence length. Moreover, in experiments the proximity-providing superconductor may dope the proximitized part of the semiconductor with additional carriers due to a difference in work functions. The corresponding difference in chemical potential causes a momentum mismatch between superconducting and normal 2DEG regions which introduces normal reflection at the superconducting-normal interface.

We focus on the scattering problem at zero energy, as we are only interested in the phase diagram. In the presence of normal reflections the scattering matrix of the left (right) normal-superconducting interface $S_{L/R}$ has the form 
\begin{align}
S_{L/R}(\phi)=e^{\pm i\phi/2\tau_{z}}Se^{\mp i\phi/2\tau_{z}};\qquad S=\begin{pmatrix}r_{e} & r_{A}\\
r_{A} & r_{h}
\end{pmatrix}
\end{align}
where $r_{e/h}$ is the normal reflection amplitude for electrons (holes). The subgap spectrum can be obtained from the condition
\begin{align}
\det(\mathbb{1}-S_{L}TS_{R}T)=0,\label{scattering_det_eq} 
\end{align}
where $T={\rm diag}(t_{e},t_{h})$ is the transmission matrix of the junction, $t_{e/h}=\exp(ik_{e/h}W)$, with $k_{e/h}$ the electron (hole) wavevector in the normal junction. The scattering amplitudes are constrained by unitarity, and can be parametrized as 
\begin{equation}
\begin{split}
r_{e/h}= & \pm r\exp(i\eta\pm i\varphi_{N}) \\
r_{A}= & (1-r^{2})^{1/2}\exp(i\eta). 
\end{split}
\label{eq:reflection_amplitudes}
\end{equation}
The phase $\eta$ depends on the superconducting gap $\Delta$, and the phase $\varphi_{N}$ depends on details of the normal reflection. After a straightforward calculation we can rewrite the condition for a subgap state as
\begin{align}
\cos(2\theta_{-}+2\eta) & =r^{2}\cos(2\theta_{+}+2\varphi_{N})+(1-r^{2})\cos\phi\label{phase_transition}
\end{align}
where we have introduced $\theta_{\pm}=(k_{e}\pm k_{h})W/2$. We can solve this equation in several limiting cases:

\begin{figure}[h]
\centering
\includegraphics[width=.48\textwidth]{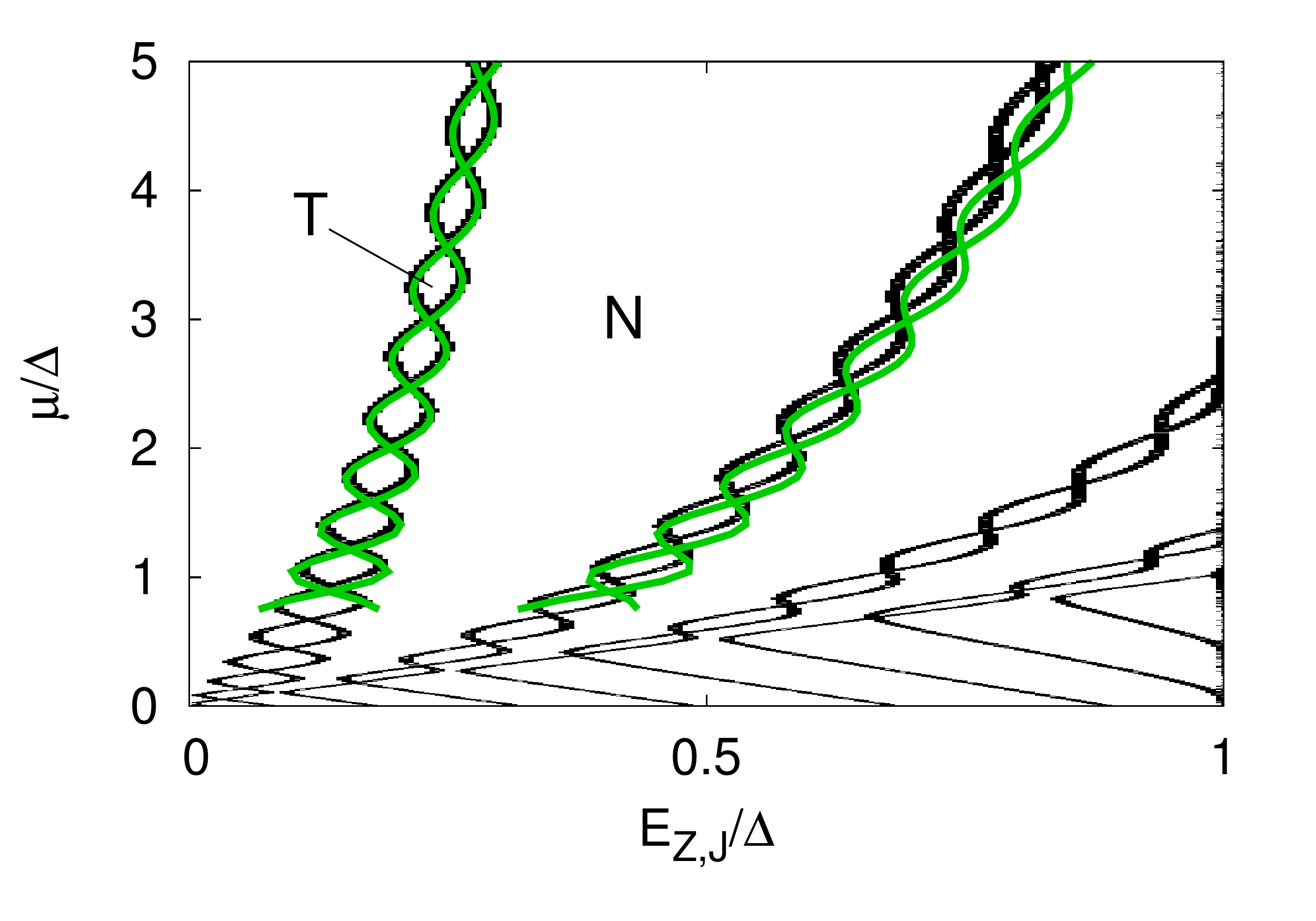}
\includegraphics[width=.48\textwidth]{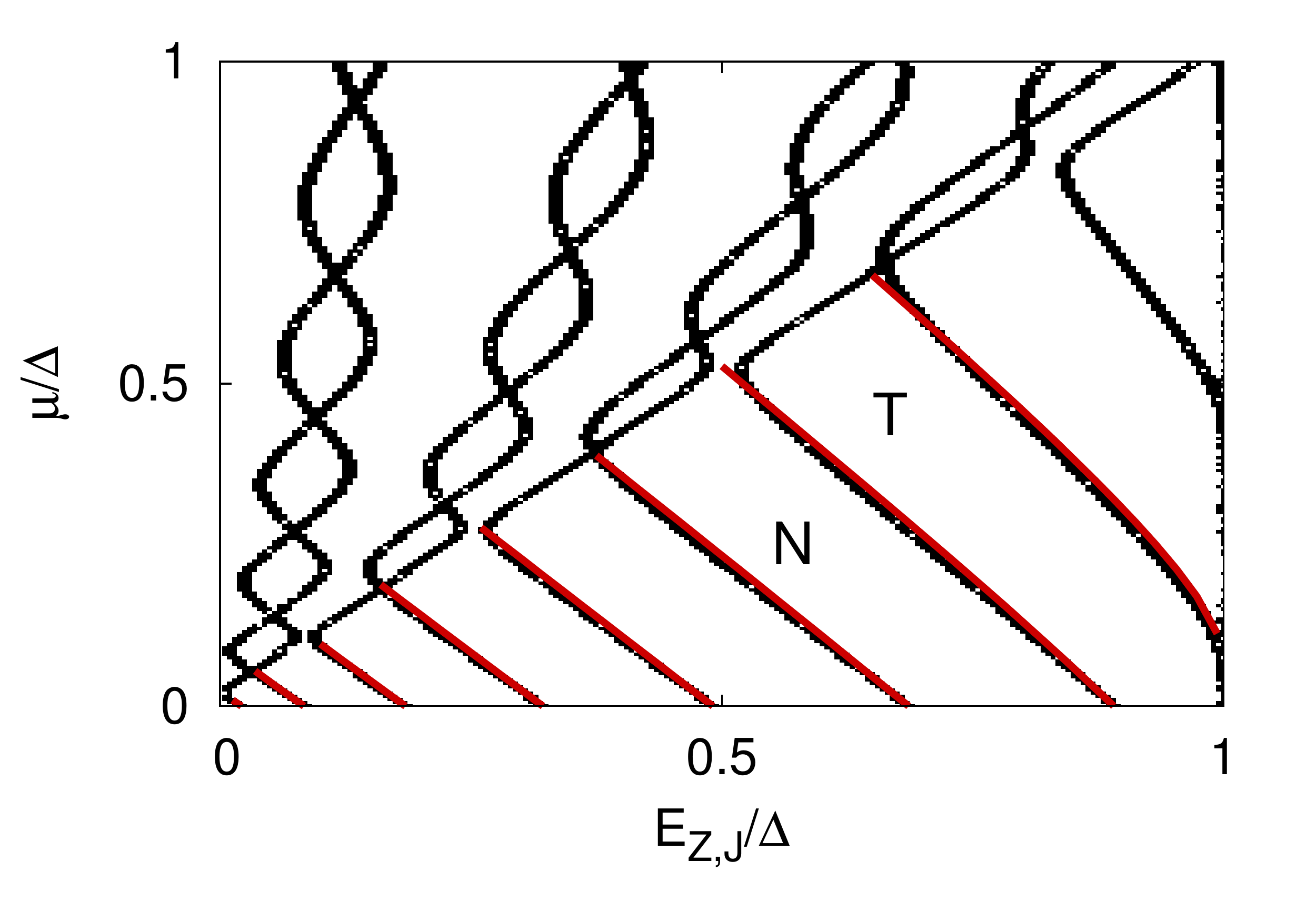}
\caption{Numerical phase diagram and analytical estimates for two limiting cases. The green lines show the solution in the limit $B\ll\mu,\Delta$ given by Eq.~(\ref{large_mu_solution}) expanded to linear order in $B/\Delta$. The normal reflection in this limit is given by $r\simeq \Delta/2\mu$ and $\varphi_N\simeq 0$. In the opposite limit $\mu\ll B<\Delta$ (red line) we use Eq.~(\ref{small_mu_solution}). Both panels show the same data for a width of $W=14\, (m\Delta)^{-1/2}$ and $\phi=0$.}
\label{fig:phase_diagram_analytics}
\end{figure}

(i) $\mu\gg E_{{\rm Z,J}}$, weak normal reflection:
In this limit we can use Eq. \eqref{eq:t_e,h} for $t_{e/h}$. The phases $\theta_{\pm}$ are then simply given by $k_{{\rm F}}W$ and $E_{{\rm Z,J}} W / v_{{\rm F}}=\phi_{B}$, respectively. The phase $\eta$ can be expanded as $\eta= \arccos\left[\left(E-E_{\rm Z,L}\right)/\Delta\right]+O(r^2)$. In the case $E_{{\rm Z,L}}=0$, the condition above for the subgap states reduces to Eq. \eqref{eq:bound_states_cond}, which was used to describe the bound states in the absence of normal reflection, with $\phi$ replaced by $\tilde{\phi}$ defined in Eq. \eqref{eq:phi_tilde} of the main text.

For $\phi=0$ and weak normal reflections $r\ll1$ the topological phase transitions are given by
\begin{align}
\phi_{B}=\pi n-\arccos(E_{{\rm Z,L}}/\Delta)\pm2r\sin\left(k_{{\rm F}}W+\varphi_{N}\right).\label{large_mu_solution} 
\end{align}
Hence in the presence of normal reflections topological phases are possible even at zero phase bias. The analytical result agrees well with numerical results shown in Fig.~\ref{fig:phase_diagram_analytics}. Similarly, for $\phi=\pi$, the phase transitions are given by 
\begin{align}
\phi_{B}=(2n+1)\frac{\pi}{2}-\arccos(E_{\rm Z,L}/\Delta)\pm2r\cos\left(k_{{\rm F}}W+\varphi_{N}\right).
\end{align}
The corrections to the scattering phase are linear in $r$ only at these two special values of $\phi$. At other values $0<\phi<\pi$ the corrections are of order $r^2$. This can be seen by comparing Figs.~\ref{fig:PD_Bmu_num}(a-c). The oscillations of the phase boundaries for $\phi=\pi/2$ shown in panel (b) vanish more rapidly with increasing $\mu$ than those in (a) and (c).

A topological phase may be accessible even in very narrow junctions where $E_{{\rm Z,J}},\Delta\ll E_{{\rm T}}$ when the Zeeman field in the lead is of order of $\Delta$. Eq.~(\ref{phase_transition}) then yields the phase boundary
\begin{align}
 E_{\rm Z,L}=\Delta\sqrt{1-r^2\sin^2(k_{\rm F}W+\varphi_N)-(1-r^2)\sin^2\phi/2}
\end{align}
The result plotted in Fig.~\ref{fig:phase_diagram_leads} as dashed lines. In suffciently short junctions when $\mu \ll E_{\rm T}$ we can set $\theta_\pm=0$. Moreover, one can show in this case that $\varphi_N\simeq 0$ when normal reflections are weak. The phase boundaries then follow the well-known dispersion of Andreev bound states in a short junction $ E_{\rm Z,L}=\Delta\sqrt{1-(1-r^2)\sin^2\phi/2}$.

(ii) $\mu\gg E_{{\rm Z,J}}$, strong normal reflection:
Normal reflection should ideally be avoided as it weakens the proximity effect and reduces the overall gap of the system. To illustrate the effect of increasing normal reflections we consider the extreme case $r\to 1$. For simplicity we also set $E_{\rm Z,L}=0$. The phase diagram then becomes independent of $\phi$ and we find
\begin{align}
\phi_B \simeq\pm(k_{\rm F}W+\varphi_{N})-\eta+n\pi
\end{align}
The phase boundaries $E_{\rm Z,J}\simeq \pm2\mu+2nE_{\rm T}+{\rm const.}$ form diamonds in the $E_{\rm Z,J}-\mu$ plane. This trend can already be seen for rather weak normal reflection in Fig.~\hyperref[fig:PD_Bmu_num]{\ref{fig:PD_Bmu_num}(d)}. Thus as normal reflection becomes stronger, the phase space decomposes into similar-sized patches of topological and trivial phase which alternate with period $k_{\rm F}W$ as a function of chemical potential. Note that when the normal reflection is strong, the system may be thought of as a wire of width $W$ weakly coupled to two superconductors. Then, the period of the oscillations corresponds to the addition of a single transverse channel to the wire.

We conclude that normal reflection is generically detrimental to topological superconductivity. Even though normal reflection increases the phase space area of the topological phase at $\phi=0$, the small patches make the topology vulnerable to potential fluctuations. Moreover tuning the topology with a phase bias becomes less efficient in the presence of normal reflection.

(iii) $\mu< E_{{\rm Z,J}}$:
The phase diagram has a qualitatively different behavior when the Zeeman energy exceeds $\mu$ as illustrated in Fig.~\ref{fig:phase_diagram_analytics}. The normal system becomes half metallic in the regime $\mu<E_{\rm Z,J}$ and thus only one spin component propagates in the normal region. Similar to case (ii), superconducting correlations inside the junction are suppressed. The phase diagram becomes largely independent of the phase difference and the induced gap is reduced. 

We focus on the eigenspace $\sigma_x=-1$ of the Hamiltonian at $k_x=0$ in Eq.~\eqref{hamil_1d}. In this subspace the hole part of the wavefunction is evanescent even in the normal region. For simplicity we assume a junction wider than the decay length $W\sqrt{2m(E_{\rm Z,J}-\mu)}\gg 1$ so that the transmission of holes through the normal part is strictly zero. The scattering matrix then only involves normal reflection of electrons whose reflection amplitude $\tilde{r}$ has unit modulus while the subgap spectrum is determined by its phase.

In this case Eq.~(\ref{scattering_det_eq}) is modified and the condition for a subgap state becomes
\begin{align}
1-\tilde{r}e^{i\theta}\tilde{r}e^{i\theta}=0
\end{align}
and thus
\begin{align}
 \tilde{r}e^{i\theta}=\pm1,
\end{align}
where $\theta=\sqrt{2m(E_{\rm Z,J}+\mu)}W$ is the phase shift of electrons traversing the normal region.

When assuming $\mu\ll E_{\rm Z,J}$ we can neglect the $\mu$ dependence of $\tilde{r}$. Calculation reveals the topological phase transitions
\begin{align}
 \mu=E_{\rm Z,J}+\frac{1}{2m}\left(\frac{\varphi_r +(2n+1)\pi/2}{W}\right)^2\label{small_mu_solution}
\end{align}
with $\varphi_r=2\arctan[1/(1+\sqrt{2E_{\rm Z,J}/\sqrt{\Delta^2-E_{\rm Z,L}^2}})]$. This result is in excellent agreement with numerical calculations as shown in Fig.~\ref{fig:phase_diagram_analytics}.

\section{Josephson current at finite temperature\label{app:Finite-temperature}}

\begin{figure}[tb]
\centering
\vspace{0.5cm}
\includegraphics[width=0.35\textwidth]{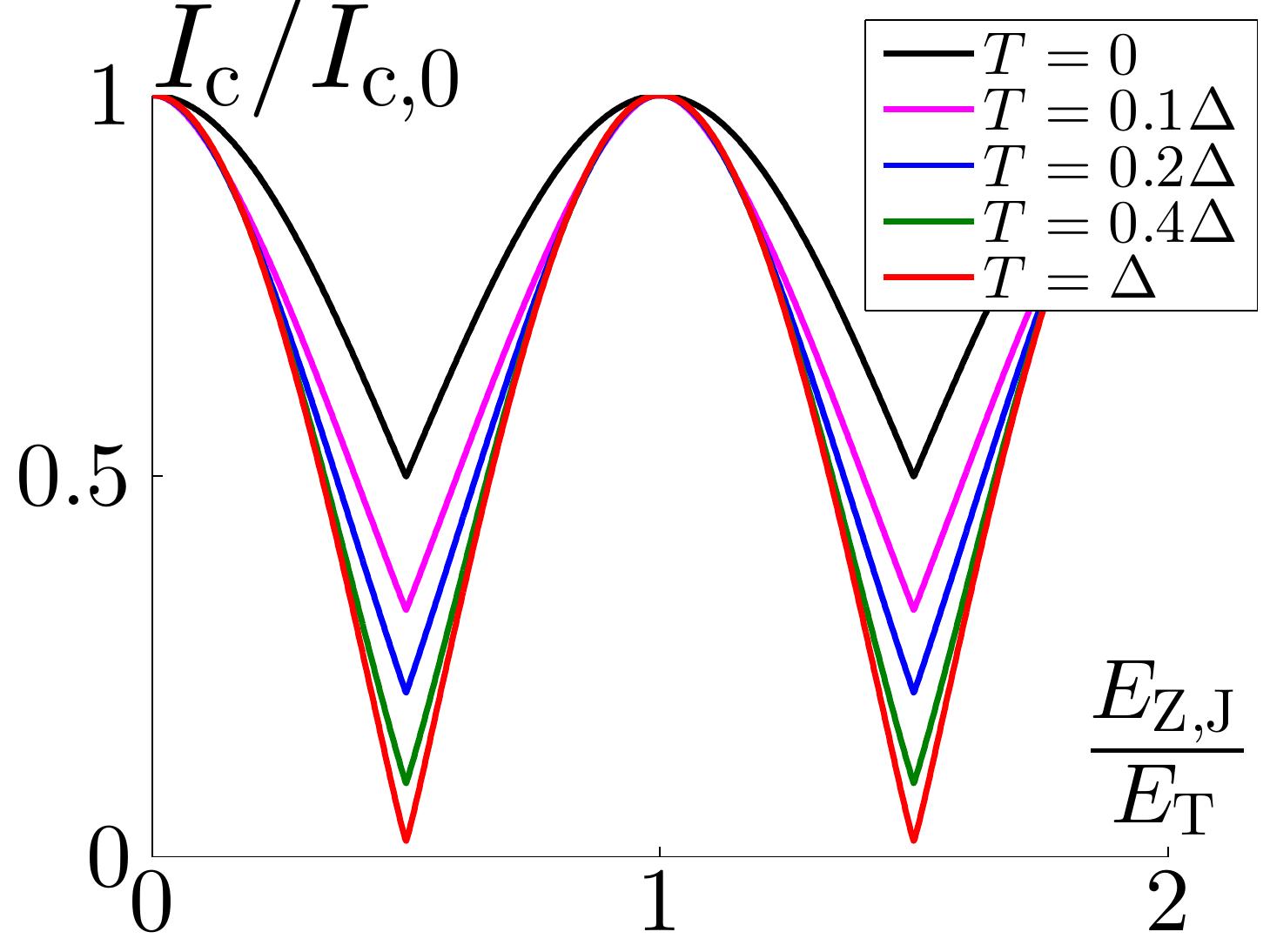}
\caption{The critical current for the $k_x=0$ mode in a narrow junction at different temperatures. As the temperature is increased the contrast of the modulations is increased with the minima at $E_{{\rm Z,J}}=\left(n+1/2\right)E_{{\rm T}}$ becoming deeper.}
\label{fig:Ic_vs_B_Tdependence}
\end{figure}

To calculate the Josephson current at finite temperature we first
calculate the many body partition function of the system. In presence
of particle-hole symmetry it is given by
\begin{equation}
Z=\prod_{n}\left(1+e^{-\beta E_{n}}\right)\left(1+e^{\beta E_{n}}\right)=4\prod_{n}{\rm \cosh}^{2}\frac{\beta E_{n}}{2},
\end{equation}
where $\beta=T^{-1}$ and the product is taken over
all the positive energy states labeled by $n$. The free energy is
then
\begin{equation}
F=-T\ln Z=-\frac{8}{\beta}\sum_{n}\ln\left(\cosh\left(\frac{\beta E_{n}}{2}\right)\right)
\end{equation}
and the Josephson current is 
\begin{equation}
I\left(\phi\right)=2e\frac{dF}{d\phi}=-8e\sum_{n}\tanh\left(\frac{\beta E_{n}}{2}\right)\frac{dE_{n}}{d\phi}.\label{eq:JCNarrow}
\end{equation}
In the high temperature limit with $\beta E_{n}\ll1$ for all the
bound states 
\begin{equation}
I\left(\phi\right)=-4e\beta\sum_{n}E_{n}\frac{dE_{n}}{d\phi}.\label{eq:JCNarrow-HighT}
\end{equation}

To show the effect of finite temperature on the critical current we calculate the current for the $k_x=0$ mode in a narrow junction with $\Delta\ll E_{{\rm T}}$ by substituting the bound state spectrum for this case given in Eq.~\eqref{eq:bs_spectrum} into Eq.~\eqref{eq:JCNarrow}. Results are plotted in Fig.~\ref{fig:Ic_vs_B_Tdependence}. It can be seen that the minima of the critical current grow deeper rapidly as the temperature is increased.

\section{Orbital effect of the in-plane magnetic field}\label{app:Orbital}

If the magnetic field below the superconducting leads is nonzero,
it is important to examine also its orbital effect. We choose a gauge
in which the vector potential is given by $\vec{A}=\left(0,0,By\right)$.
An electron tunneling between the 2DEG and the superconducting leads
acquires a position dependent phase $t_{\perp}\sim e^{iA_{z}\tilde{d}}=e^{iB\tilde{d}y}$,
where we denote by $\tilde{d}=d+\lambda_{L}$ the sum of the distance
between the 2DEG and the superconductors and the London penetration
depth. The induced order parameter therefore varies in space as $\Delta\left(y\right)=\Delta e^{iqy}$,
where $q=2B\tilde{d}$. If the correlation length of the induced pairing
is smaller than the width of the superconductors, $W_{{\rm SC}}$,
then the effective superconducting pairing will be 
\begin{equation}
\bar{\Delta}=\frac{1}{W_{{\rm SC}}}\int_{0}^{W_{{\rm SC}}}\Delta e^{iqy}=\Delta e^{i\frac{qW_{{\rm SC}}}{2}}{\rm sinc}\left(\frac{qW_{{\rm SC}}}{2}\right).
\end{equation}
Hence for values of $B$ equal to an integer multiple of $\pi/\left(\tilde{d}W_{{\rm SC}}\right)$
the superconducting gap will close, resulting, in particular, in a
vanishing critical current.

\section{Numerical calculations}
\subsection{Scattering matrix formalism}\label{app:ScatteringMatrix}

To evaluate the phase diagram and the gap of the Hamiltonian in Eq.~\eqref{hamil} we employ a numerical method based on a scattering matrix approach \cite{Bardarson2007,Brouwer2011}. From the scattering matrix $S$ we can obtain the bound state energies $\epsilon$ by solving $\det[\mathbb{1}-S(\epsilon)]=0$. For the phase diagram in Fig.~\ref{fig:PD_Bmu_num} we plot the lowest positive energy eigenvalue at $k_x=0$ in a color scale, where energies below (above) $\epsilon=0.01$ are plotted in black (blue). The gap in Fig.~\ref{fig:gap} is found by minimizing the lowest energy eigenvalue over all $k_x$.

\subsection{Tight binding model\label{app:Tight-binding}}

The tight-binding version of the Hamiltonian we use for various calculations throughout the manuscript is given by
\begin{widetext}
\begin{align}
H_{\rm TB}&=H_0+H_{\rm SOC}+H_Z+H_\Delta\\
H_{0} & =-\mu \underset{sij}{\sum}c_{i,j,s}^{\dagger}c_{i,j,s}
-t\underset{\braket{ij,i'j'}s}{\sum}\left[c_{i,j,s}^{\dagger}c_{i',j',s}+{\rm h.c.}\right]\\
H_{\rm SOC} & =i\alpha\underset{s,s'}{\sum}\left[
\underset{i=1}{\overset{L-1}{\sum}}\ \underset{j=1}{\overset{2W_{SC}+W}{\sum}}
c_{i+1,j,s}^{\dagger}\sigma_{y}^{s,s'}c_{i,j,s'}
-\underset{i=1}{\overset{L}{\sum}}\underset{j=1}{\overset{2W_{SC}+W-1}{\sum}}
c_{i,j+1,s}^{\dagger}\sigma_{x}^{s,s'}c_{i,j,s'}
-{\rm h.c.}\right]\\
H_{Z} & =\underset{iss'}{\sum}
\left(
E_{{\rm Z,J}}\underset{j=W_{SC}+1}{\overset{W_{SC}+W}{\sum}}
+E_{{\rm Z,L}}\underset{j=1}{\overset{W_{SC}}{\sum}}
+E_{{\rm Z,L}}\underset{j=W_{SC}+W+1}{\overset{2W_{SC}+W}{\sum}}\right)
\left(c_{i,j,s}^{\dagger}\sigma_{x}^{s,s'}c_{i,j,s'}+{\rm h.c.}\right)\\
H_{\Delta} & =\Delta e^{-i\phi/2}\underset{i}{\sum}\underset{j=1}{\overset{W_{SC}}{\sum}}c_{i,j,\uparrow}c_{i,j,\downarrow}+\Delta e^{i\phi/2}\underset{i}{\sum}\underset{j=W_{SC}+W+1}{\overset{2W_{SC}+W}{\sum}}c_{i,j,\uparrow}c_{i,j,\downarrow}+{\rm h.c.},
\end{align}
\end{widetext}
where $c_{i,j,s}$ is the annihilation operator of an electron spin $s$ on site $(i,j)$ with $1\leq j\leq 2W_{SC}+W$ and $1\leq i\leq L$ and $\braket{.,.}$ denotes nearest neighbors. The hopping and spin-orbit coupling strength are denoted by $t,\alpha$ respectively. Proximity induced pairing strength $\Delta$ is nonzero only in the leads $1\leq j\leq W_{SC}$ and $W_{SC}+W<j\leq2W_{SC}+W$. The Zeeman field along the $x$-direction has strength $E_{\rm Z,L}$ ($E_{\rm Z,J}$) in the leads (in the junction). 
This model has been used to calculate the local density of states in Fig.~\ref{fig:ldos}.

To describe an infinitely long junction we assume $L\to \infty$ and perform a partial Fourier transform $c_{j,k,s}=\sum_je^{iki'}c_{j,i',s}$. The resulting Hamiltonian $H_{\rm TB}(k)$ is used to calculate the BDI phase diagram in Fig.~\ref{fig:BDI_PD}, the gap in the system when the effective time-reversal symmetry is broken in Fig.~\ref{fig:BDI_gap}, the spectrum across the topological phase transition of class D in Fig.~\ref{fig:spectrum_vs_kx} and to demonstrate the first order phase transition accompanied by a minimum in the critical current in Fig.~\hyperref[fig:Ic]{\ref{fig:Ic}(b)}.

\bibliography{calculations,ref}

\begin{thebibliography}{34}
\expandafter\ifx\csname natexlab\endcsname\relax\def\natexlab#1{#1}\fi
\expandafter\ifx\csname bibnamefont\endcsname\relax
  \def\bibnamefont#1{#1}\fi
\expandafter\ifx\csname bibfnamefont\endcsname\relax
  \def\bibfnamefont#1{#1}\fi
\expandafter\ifx\csname citenamefont\endcsname\relax
  \def\citenamefont#1{#1}\fi
\expandafter\ifx\csname url\endcsname\relax
  \def\url#1{\texttt{#1}}\fi
\expandafter\ifx\csname urlprefix\endcsname\relax\def\urlprefix{URL }\fi
\providecommand{\bibinfo}[2]{#2}
\providecommand{\eprint}[2][]{\url{#2}}

\bibitem[{\citenamefont{Alicea}(2012)}]{Alicea2012}
\bibinfo{author}{\bibfnamefont{J.}~\bibnamefont{Alicea}},
  \bibinfo{journal}{Reports on Progress in Physics}
  \textbf{\bibinfo{volume}{75}}, \bibinfo{pages}{076501}
  (\bibinfo{year}{2012}),
  \urlprefix\url{http://stacks.iop.org/0034-4885/75/i=7/a=076501}.

\bibitem[{\citenamefont{Beenakker}(2013)}]{Beenakker2013}
\bibinfo{author}{\bibfnamefont{C.}~\bibnamefont{Beenakker}},
  \bibinfo{journal}{Annual Review of Condensed Matter Physics}
  \textbf{\bibinfo{volume}{4}}, \bibinfo{pages}{113} (\bibinfo{year}{2013}),
  \urlprefix\url{http://dx.doi.org/10.1146/annurev-conmatphys-030212-184337}.

\bibitem[{\citenamefont{Mourik et~al.}(2012)\citenamefont{Mourik, Zuo, Frolov,
  Plissard, Bakkers, and Kouwenhoven}}]{Mourik2012}
\bibinfo{author}{\bibfnamefont{V.}~\bibnamefont{Mourik}},
  \bibinfo{author}{\bibfnamefont{K.}~\bibnamefont{Zuo}},
  \bibinfo{author}{\bibfnamefont{S.~M.} \bibnamefont{Frolov}},
  \bibinfo{author}{\bibfnamefont{S.~R.} \bibnamefont{Plissard}},
  \bibinfo{author}{\bibfnamefont{E.~P. A.~M.} \bibnamefont{Bakkers}},
  \bibnamefont{and} \bibinfo{author}{\bibfnamefont{L.~P.}
  \bibnamefont{Kouwenhoven}}, \bibinfo{journal}{Science}
  \textbf{\bibinfo{volume}{336}}, \bibinfo{pages}{1003} (\bibinfo{year}{2012}),
  ISSN \bibinfo{issn}{0036-8075},
  \urlprefix\url{http://science.sciencemag.org/content/336/6084/1003}.

\bibitem[{\citenamefont{Das et~al.}(2012)\citenamefont{Das, Ronen, Most, Oreg,
  Heiblum, and Shtrikman}}]{Das2012}
\bibinfo{author}{\bibfnamefont{A.}~\bibnamefont{Das}},
  \bibinfo{author}{\bibfnamefont{Y.}~\bibnamefont{Ronen}},
  \bibinfo{author}{\bibfnamefont{Y.}~\bibnamefont{Most}},
  \bibinfo{author}{\bibfnamefont{Y.}~\bibnamefont{Oreg}},
  \bibinfo{author}{\bibfnamefont{M.}~\bibnamefont{Heiblum}}, \bibnamefont{and}
  \bibinfo{author}{\bibfnamefont{H.}~\bibnamefont{Shtrikman}},
  \bibinfo{journal}{Nat. Phys.} \textbf{\bibinfo{volume}{8}},
  \bibinfo{pages}{887} (\bibinfo{year}{2012}), ISSN \bibinfo{issn}{1745-2473},
  \urlprefix\url{http://www.nature.com/nphys/journal/v8/n12/full/nphys2479.html}.

\bibitem[{\citenamefont{Nadj-Perge et~al.}(2014)\citenamefont{Nadj-Perge,
  Drozdov, Li, Chen, Jeon, Seo, MacDonald, Bernevig, and
  Yazdani}}]{Nadj-Perge2014}
\bibinfo{author}{\bibfnamefont{S.}~\bibnamefont{Nadj-Perge}},
  \bibinfo{author}{\bibfnamefont{I.~K.} \bibnamefont{Drozdov}},
  \bibinfo{author}{\bibfnamefont{J.}~\bibnamefont{Li}},
  \bibinfo{author}{\bibfnamefont{H.}~\bibnamefont{Chen}},
  \bibinfo{author}{\bibfnamefont{S.}~\bibnamefont{Jeon}},
  \bibinfo{author}{\bibfnamefont{J.}~\bibnamefont{Seo}},
  \bibinfo{author}{\bibfnamefont{A.~H.} \bibnamefont{MacDonald}},
  \bibinfo{author}{\bibfnamefont{B.~A.} \bibnamefont{Bernevig}},
  \bibnamefont{and} \bibinfo{author}{\bibfnamefont{A.}~\bibnamefont{Yazdani}},
  \bibinfo{journal}{Science} \textbf{\bibinfo{volume}{346}},
  \bibinfo{pages}{602} (\bibinfo{year}{2014}), ISSN \bibinfo{issn}{0036-8075},
  \urlprefix\url{http://science.sciencemag.org/content/346/6209/602}.

\bibitem[{\citenamefont{Albrecht et~al.}(2016)\citenamefont{Albrecht,
  Higginbotham, Madsen, Kuemmeth, Jespersen, Nyg{\aa}rd, Krogstrup, and
  Marcus}}]{Albrecht2016}
\bibinfo{author}{\bibfnamefont{S.~M.} \bibnamefont{Albrecht}},
  \bibinfo{author}{\bibfnamefont{A.~P.} \bibnamefont{Higginbotham}},
  \bibinfo{author}{\bibfnamefont{M.}~\bibnamefont{Madsen}},
  \bibinfo{author}{\bibfnamefont{F.}~\bibnamefont{Kuemmeth}},
  \bibinfo{author}{\bibfnamefont{T.~S.} \bibnamefont{Jespersen}},
  \bibinfo{author}{\bibfnamefont{J.}~\bibnamefont{Nyg{\aa}rd}},
  \bibinfo{author}{\bibfnamefont{P.}~\bibnamefont{Krogstrup}},
  \bibnamefont{and} \bibinfo{author}{\bibfnamefont{C.~M.}
  \bibnamefont{Marcus}}, \bibinfo{journal}{Nature}
  \textbf{\bibinfo{volume}{531}}, \bibinfo{pages}{206} (\bibinfo{year}{2016}),
  ISSN \bibinfo{issn}{0028-0836},
  \urlprefix\url{http://dx.doi.org/10.1038/nature17162}.

\bibitem[{\citenamefont{Hart et~al.}(2015)\citenamefont{Hart, Ren, Kosowsky,
  Ben-Shach, Leubner, Br{\"u}ne, Buhmann, Molenkamp, Halperin, and
  Yacoby}}]{hart2015controlled}
\bibinfo{author}{\bibfnamefont{S.}~\bibnamefont{Hart}},
  \bibinfo{author}{\bibfnamefont{H.}~\bibnamefont{Ren}},
  \bibinfo{author}{\bibfnamefont{M.}~\bibnamefont{Kosowsky}},
  \bibinfo{author}{\bibfnamefont{G.}~\bibnamefont{Ben-Shach}},
  \bibinfo{author}{\bibfnamefont{P.}~\bibnamefont{Leubner}},
  \bibinfo{author}{\bibfnamefont{C.}~\bibnamefont{Br{\"u}ne}},
  \bibinfo{author}{\bibfnamefont{H.}~\bibnamefont{Buhmann}},
  \bibinfo{author}{\bibfnamefont{L.~W.} \bibnamefont{Molenkamp}},
  \bibinfo{author}{\bibfnamefont{B.~I.} \bibnamefont{Halperin}},
  \bibnamefont{and} \bibinfo{author}{\bibfnamefont{A.}~\bibnamefont{Yacoby}},
  \bibinfo{journal}{arXiv:1509.02940}  (\bibinfo{year}{2015}).

\bibitem[{\citenamefont{Wan et~al.}(2015)\citenamefont{Wan, Kazakov, Manfra,
  Pfeiffer, West, and Rokhinson}}]{Wan2015}
\bibinfo{author}{\bibfnamefont{Z.}~\bibnamefont{Wan}},
  \bibinfo{author}{\bibfnamefont{A.}~\bibnamefont{Kazakov}},
  \bibinfo{author}{\bibfnamefont{M.~J.} \bibnamefont{Manfra}},
  \bibinfo{author}{\bibfnamefont{L.~N.} \bibnamefont{Pfeiffer}},
  \bibinfo{author}{\bibfnamefont{K.~W.} \bibnamefont{West}}, \bibnamefont{and}
  \bibinfo{author}{\bibfnamefont{L.~P.} \bibnamefont{Rokhinson}},
  \bibinfo{journal}{Nat Commun} \textbf{\bibinfo{volume}{6}},
  (\bibinfo{year}{2015}), \urlprefix\url{http://dx.doi.org/10.1038/ncomms8426}.

\bibitem[{\citenamefont{Kjaergaard et~al.}(2016)\citenamefont{Kjaergaard,
  Suominen, Nowak, Akhmerov, Shabani, Palmstr{\o}m, Nichele, and
  Marcus}}]{Kjaergaard2016}
\bibinfo{author}{\bibfnamefont{M.}~\bibnamefont{Kjaergaard}},
  \bibinfo{author}{\bibfnamefont{H.~J.} \bibnamefont{Suominen}},
  \bibinfo{author}{\bibfnamefont{M.~P.} \bibnamefont{Nowak}},
  \bibinfo{author}{\bibfnamefont{A.~R.} \bibnamefont{Akhmerov}},
  \bibinfo{author}{\bibfnamefont{J.}~\bibnamefont{Shabani}},
  \bibinfo{author}{\bibfnamefont{C.~J.} \bibnamefont{Palmstr{\o}m}},
  \bibinfo{author}{\bibfnamefont{F.}~\bibnamefont{Nichele}}, \bibnamefont{and}
  \bibinfo{author}{\bibfnamefont{C.~M.} \bibnamefont{Marcus}}
  (\bibinfo{year}{2016}), \eprint{arXiv:1607.04164}.

\bibitem[{\citenamefont{Hyart et~al.}(2013)\citenamefont{Hyart, van Heck,
  Fulga, Burrello, Akhmerov, and Beenakker}}]{Hyart2013}
\bibinfo{author}{\bibfnamefont{T.}~\bibnamefont{Hyart}},
  \bibinfo{author}{\bibfnamefont{B.}~\bibnamefont{van Heck}},
  \bibinfo{author}{\bibfnamefont{I.~C.} \bibnamefont{Fulga}},
  \bibinfo{author}{\bibfnamefont{M.}~\bibnamefont{Burrello}},
  \bibinfo{author}{\bibfnamefont{A.~R.} \bibnamefont{Akhmerov}},
  \bibnamefont{and} \bibinfo{author}{\bibfnamefont{C.~W.~J.}
  \bibnamefont{Beenakker}}, \bibinfo{journal}{Phys. Rev. B}
  \textbf{\bibinfo{volume}{88}}, \bibinfo{pages}{035121}
  (\bibinfo{year}{2013}),
  \urlprefix\url{http://link.aps.org/doi/10.1103/PhysRevB.88.035121}.

\bibitem[{\citenamefont{Li et~al.}(2016)\citenamefont{Li, Neupert, Bernevig,
  and Yazdani}}]{Li2016}
\bibinfo{author}{\bibfnamefont{J.}~\bibnamefont{Li}},
  \bibinfo{author}{\bibfnamefont{T.}~\bibnamefont{Neupert}},
  \bibinfo{author}{\bibfnamefont{B.~A.} \bibnamefont{Bernevig}},
  \bibnamefont{and} \bibinfo{author}{\bibfnamefont{A.}~\bibnamefont{Yazdani}},
  \bibinfo{journal}{Nature Communications} \textbf{\bibinfo{volume}{7}},
  \bibinfo{pages}{10395} (\bibinfo{year}{2016}),
  \urlprefix\url{http://dx.doi.org/10.1038/ncomms10395}.

\bibitem[{\citenamefont{Shabani et~al.}(2016)\citenamefont{Shabani, Kjaergaard,
  Suominen, Kim, Nichele, Pakrouski, Stankevic, Lutchyn, Krogstrup,
  Feidenhans'l et~al.}}]{Shabani2016}
\bibinfo{author}{\bibfnamefont{J.}~\bibnamefont{Shabani}},
  \bibinfo{author}{\bibfnamefont{M.}~\bibnamefont{Kjaergaard}},
  \bibinfo{author}{\bibfnamefont{H.~J.} \bibnamefont{Suominen}},
  \bibinfo{author}{\bibfnamefont{Y.}~\bibnamefont{Kim}},
  \bibinfo{author}{\bibfnamefont{F.}~\bibnamefont{Nichele}},
  \bibinfo{author}{\bibfnamefont{K.}~\bibnamefont{Pakrouski}},
  \bibinfo{author}{\bibfnamefont{T.}~\bibnamefont{Stankevic}},
  \bibinfo{author}{\bibfnamefont{R.~M.} \bibnamefont{Lutchyn}},
  \bibinfo{author}{\bibfnamefont{P.}~\bibnamefont{Krogstrup}},
  \bibinfo{author}{\bibfnamefont{R.}~\bibnamefont{Feidenhans'l}},
  \bibnamefont{et~al.}, \bibinfo{journal}{Phys. Rev. B}
  \textbf{\bibinfo{volume}{93}}, \bibinfo{pages}{155402}
  (\bibinfo{year}{2016}),
  \urlprefix\url{http://link.aps.org/doi/10.1103/PhysRevB.93.155402}.

\bibitem[{\citenamefont{Vuik et~al.}(2016)\citenamefont{Vuik, Eeltink,
  Akhmerov, and Wimmer}}]{Vuik2016}
\bibinfo{author}{\bibfnamefont{A.}~\bibnamefont{Vuik}},
  \bibinfo{author}{\bibfnamefont{D.}~\bibnamefont{Eeltink}},
  \bibinfo{author}{\bibfnamefont{A.~R.} \bibnamefont{Akhmerov}},
  \bibnamefont{and} \bibinfo{author}{\bibfnamefont{M.}~\bibnamefont{Wimmer}},
  \bibinfo{journal}{New Journal of Physics} \textbf{\bibinfo{volume}{18}},
  \bibinfo{pages}{033013} (\bibinfo{year}{2016}),
  \urlprefix\url{http://stacks.iop.org/1367-2630/18/i=3/a=033013}.

\bibitem[{\citenamefont{Altland and Zirnbauer}(1997)}]{AltlandZirnbauer1997}
\bibinfo{author}{\bibfnamefont{A.}~\bibnamefont{Altland}} \bibnamefont{and}
  \bibinfo{author}{\bibfnamefont{M.~R.} \bibnamefont{Zirnbauer}},
  \bibinfo{journal}{Physical Review B} \textbf{\bibinfo{volume}{55}},
  \bibinfo{pages}{1142} (\bibinfo{year}{1997}),
  \urlprefix\url{http://link.aps.org/doi/10.1103/PhysRevB.55.1142}.

\bibitem[{\citenamefont{Mizushima and Sato}(2013)}]{Mizushima2013}
\bibinfo{author}{\bibfnamefont{T.}~\bibnamefont{Mizushima}} \bibnamefont{and}
  \bibinfo{author}{\bibfnamefont{M.}~\bibnamefont{Sato}}, \bibinfo{journal}{New
  Journal of Physics} \textbf{\bibinfo{volume}{15}}, \bibinfo{pages}{075010}
  (\bibinfo{year}{2013}),
  \urlprefix\url{http://stacks.iop.org/1367-2630/15/i=7/a=075010}.

\bibitem[{\citenamefont{Ryazanov et~al.}(2001)\citenamefont{Ryazanov, Oboznov,
  Rusanov, Veretennikov, Golubov, and Aarts}}]{Ryazanov2001}
\bibinfo{author}{\bibfnamefont{V.~V.} \bibnamefont{Ryazanov}},
  \bibinfo{author}{\bibfnamefont{V.~A.} \bibnamefont{Oboznov}},
  \bibinfo{author}{\bibfnamefont{A.~Y.} \bibnamefont{Rusanov}},
  \bibinfo{author}{\bibfnamefont{A.~V.} \bibnamefont{Veretennikov}},
  \bibinfo{author}{\bibfnamefont{A.~A.} \bibnamefont{Golubov}},
  \bibnamefont{and} \bibinfo{author}{\bibfnamefont{J.}~\bibnamefont{Aarts}},
  \bibinfo{journal}{Phys. Rev. Lett.} \textbf{\bibinfo{volume}{86}},
  \bibinfo{pages}{2427} (\bibinfo{year}{2001}),
  \urlprefix\url{http://link.aps.org/doi/10.1103/PhysRevLett.86.2427}.

\bibitem[{\citenamefont{Kontos et~al.}(2002)\citenamefont{Kontos, Aprili,
  Lesueur, Gen\^et, Stephanidis, and Boursier}}]{Kontos2002}
\bibinfo{author}{\bibfnamefont{T.}~\bibnamefont{Kontos}},
  \bibinfo{author}{\bibfnamefont{M.}~\bibnamefont{Aprili}},
  \bibinfo{author}{\bibfnamefont{J.}~\bibnamefont{Lesueur}},
  \bibinfo{author}{\bibfnamefont{F.}~\bibnamefont{Gen\^et}},
  \bibinfo{author}{\bibfnamefont{B.}~\bibnamefont{Stephanidis}},
  \bibnamefont{and} \bibinfo{author}{\bibfnamefont{R.}~\bibnamefont{Boursier}},
  \bibinfo{journal}{Phys. Rev. Lett.} \textbf{\bibinfo{volume}{89}},
  \bibinfo{pages}{137007} (\bibinfo{year}{2002}),
  \urlprefix\url{http://link.aps.org/doi/10.1103/PhysRevLett.89.137007}.

\bibitem[{\citenamefont{Frolov et~al.}(2004)\citenamefont{Frolov,
  Van~Harlingen, Oboznov, Bolginov, and Ryazanov}}]{Frolov2004}
\bibinfo{author}{\bibfnamefont{S.~M.} \bibnamefont{Frolov}},
  \bibinfo{author}{\bibfnamefont{D.~J.} \bibnamefont{Van~Harlingen}},
  \bibinfo{author}{\bibfnamefont{V.~A.} \bibnamefont{Oboznov}},
  \bibinfo{author}{\bibfnamefont{V.~V.} \bibnamefont{Bolginov}},
  \bibnamefont{and} \bibinfo{author}{\bibfnamefont{V.~V.}
  \bibnamefont{Ryazanov}}, \bibinfo{journal}{Phys. Rev. B}
  \textbf{\bibinfo{volume}{70}}, \bibinfo{pages}{144505}
  (\bibinfo{year}{2004}),
  \urlprefix\url{http://link.aps.org/doi/10.1103/PhysRevB.70.144505}.

\bibitem[{TBP()}]{TBParams}
\bibinfo{note}{See Appendix \ref{app:Tight-binding} for details of the model.
  The parameters used are: $W=5$, $W_{SC}=10$, $t=1$, $\alpha=0.1$, $\mu=-2.4$,
  $\Delta=0.3$. The phase diagram is obtained by calculating the topological
  invariant for class D, $Q={\rm sign}\left[{\rm
  Pf}\left(H_{k=\pi}\tau_x\right) / {\rm Pf}\left(H_{k=0}\tau_x\right)\right]$
  \cite{Tewari2012}. To obtain the phase difference which minimizes the free
  energy of the system, as well as the critical current across the junction, a
  summation over the contribution of all $k_{x}$ momenta to the free energy was
  performed.}

\bibitem[{\citenamefont{Kitaev}(2009)}]{Kitaev2009}
\bibinfo{author}{\bibfnamefont{A.~Y.} \bibnamefont{Kitaev}},
  \bibinfo{journal}{AIP Conf. Proc.} \textbf{\bibinfo{volume}{1134}},
  \bibinfo{pages}{22} (\bibinfo{year}{2009}).

\bibitem[{\citenamefont{Schnyder et~al.}(2009)\citenamefont{Schnyder, Ryu,
  Furusaki, and Ludwig}}]{Schnyder2009}
\bibinfo{author}{\bibfnamefont{A.~P.} \bibnamefont{Schnyder}},
  \bibinfo{author}{\bibfnamefont{S.}~\bibnamefont{Ryu}},
  \bibinfo{author}{\bibfnamefont{A.}~\bibnamefont{Furusaki}}, \bibnamefont{and}
  \bibinfo{author}{\bibfnamefont{A.~W.~W.} \bibnamefont{Ludwig}},
  \bibinfo{journal}{{AIP} Conference Proceedings}
  \textbf{\bibinfo{volume}{1134}}, \bibinfo{pages}{10} (\bibinfo{year}{2009}),
  ISSN \bibinfo{issn}{{0094243X}},
  \urlprefix\url{http://proceedings.aip.org/resource/2/apcpcs/1134/1/10_1?isAuthorized=no}.

\bibitem[{foo()}]{foot1}
\bibinfo{note}{In the presence of interactions, the $\mathbb{Z}$ classification
  is modified to $\mathbb{Z}_8$ \cite{Fidkowski2010}}.

\bibitem[{\citenamefont{Fidkowski and Kitaev}(2010)}]{Fidkowski2010}
\bibinfo{author}{\bibfnamefont{L.}~\bibnamefont{Fidkowski}} \bibnamefont{and}
  \bibinfo{author}{\bibfnamefont{A.}~\bibnamefont{Kitaev}},
  \bibinfo{journal}{Phys. Rev. B} \textbf{\bibinfo{volume}{81}},
  \bibinfo{pages}{134509} (\bibinfo{year}{2010}),
  \urlprefix\url{http://link.aps.org/doi/10.1103/PhysRevB.81.134509}.

\bibitem[{\citenamefont{Kitaev}(2001)}]{Kitaev2001}
\bibinfo{author}{\bibfnamefont{A.~Y.} \bibnamefont{Kitaev}},
  \bibinfo{journal}{Physics-Uspekhi} \textbf{\bibinfo{volume}{44}},
  \bibinfo{pages}{131} (\bibinfo{year}{2001}),
  \urlprefix\url{http://stacks.iop.org/1063-7869/44/i=10S/a=S29}.

\bibitem[{\citenamefont{Dolcini et~al.}(2015)\citenamefont{Dolcini, Houzet, and
  Meyer}}]{Dolcini2016}
\bibinfo{author}{\bibfnamefont{F.}~\bibnamefont{Dolcini}},
  \bibinfo{author}{\bibfnamefont{M.}~\bibnamefont{Houzet}}, \bibnamefont{and}
  \bibinfo{author}{\bibfnamefont{J.~S.} \bibnamefont{Meyer}},
  \bibinfo{journal}{Phys. Rev. B} \textbf{\bibinfo{volume}{92}},
  \bibinfo{pages}{035428} (\bibinfo{year}{2015}),
  \urlprefix\url{http://link.aps.org/doi/10.1103/PhysRevB.92.035428}.

\bibitem[{\citenamefont{Beenakker}(1991)}]{Beenakker1991}
\bibinfo{author}{\bibfnamefont{C.~W.~J.} \bibnamefont{Beenakker}},
  \bibinfo{journal}{Phys. Rev. Lett.} \textbf{\bibinfo{volume}{67}},
  \bibinfo{pages}{3836} (\bibinfo{year}{1991}),
  \urlprefix\url{http://link.aps.org/doi/10.1103/PhysRevLett.67.3836}.

\bibitem[{\citenamefont{Tewari and Sau}(2012)}]{Tewari2012}
\bibinfo{author}{\bibfnamefont{S.}~\bibnamefont{Tewari}} \bibnamefont{and}
  \bibinfo{author}{\bibfnamefont{J.~D.} \bibnamefont{Sau}},
  \bibinfo{journal}{Phys. Rev. Lett.} \textbf{\bibinfo{volume}{109}},
  \bibinfo{pages}{150408} (\bibinfo{year}{2012}),
  \urlprefix\url{http://link.aps.org/doi/10.1103/PhysRevLett.109.150408}.

\bibitem[{\citenamefont{Yokoyama and Nazarov}(2014)}]{Nazarov2014}
\bibinfo{author}{\bibfnamefont{T.}~\bibnamefont{Yokoyama}} \bibnamefont{and}
  \bibinfo{author}{\bibfnamefont{Y.~V.} \bibnamefont{Nazarov}},
  \bibinfo{journal}{EPL (Europhysics Letters)} \textbf{\bibinfo{volume}{108}},
  \bibinfo{pages}{47009} (\bibinfo{year}{2014}),
  \urlprefix\url{http://stacks.iop.org/0295-5075/108/i=4/a=47009}.

\bibitem[{\citenamefont{Ishii}(1970)}]{ishii1970josephson}
\bibinfo{author}{\bibfnamefont{C.}~\bibnamefont{Ishii}},
  \bibinfo{journal}{Progress of theoretical Physics}
  \textbf{\bibinfo{volume}{44}}, \bibinfo{pages}{1525} (\bibinfo{year}{1970}).

\bibitem[{\citenamefont{Zagoskin}(2014)}]{zagoskin2014book}
\bibinfo{author}{\bibfnamefont{A.~M.} \bibnamefont{Zagoskin}},
  \emph{\bibinfo{title}{Quantum theory of many-body systems}}
  (\bibinfo{publisher}{Springer International Publishing},
  \bibinfo{year}{2014}).

\bibitem[{\citenamefont{Hell et~al.}(2016)\citenamefont{Hell, Leijnse, and
  Flensberg}}]{Hell2016}
\bibinfo{author}{\bibfnamefont{M.}~\bibnamefont{Hell}},
  \bibinfo{author}{\bibfnamefont{M.}~\bibnamefont{Leijnse}}, \bibnamefont{and}
  \bibinfo{author}{\bibfnamefont{K.}~\bibnamefont{Flensberg}},
  \bibinfo{journal}{arXiv:1608.08769}  (\bibinfo{year}{2016}).

\bibitem[{\citenamefont{Sticlet et~al.}(2016)\citenamefont{Sticlet, Nijholt,
  and Akhmerov}}]{Sticlet2016}
\bibinfo{author}{\bibfnamefont{D.}~\bibnamefont{Sticlet}},
  \bibinfo{author}{\bibfnamefont{B.}~\bibnamefont{Nijholt}}, \bibnamefont{and}
  \bibinfo{author}{\bibfnamefont{A.}~\bibnamefont{Akhmerov}},
  \bibinfo{journal}{arXiv:1609.00637}  (\bibinfo{year}{2016}).

\bibitem[{\citenamefont{Bardarson et~al.}(2007)\citenamefont{Bardarson,
  Tworzyd\l{}o, Brouwer, and Beenakker}}]{Bardarson2007}
\bibinfo{author}{\bibfnamefont{J.~H.} \bibnamefont{Bardarson}},
  \bibinfo{author}{\bibfnamefont{J.}~\bibnamefont{Tworzyd\l{}o}},
  \bibinfo{author}{\bibfnamefont{P.~W.} \bibnamefont{Brouwer}},
  \bibnamefont{and} \bibinfo{author}{\bibfnamefont{C.~W.~J.}
  \bibnamefont{Beenakker}}, \bibinfo{journal}{Phys. Rev. Lett.}
  \textbf{\bibinfo{volume}{99}}, \bibinfo{pages}{106801}
  (\bibinfo{year}{2007}),
  \urlprefix\url{http://link.aps.org/doi/10.1103/PhysRevLett.99.106801}.

\bibitem[{\citenamefont{Brouwer et~al.}(2011)\citenamefont{Brouwer, Duckheim,
  Romito, and von Oppen}}]{Brouwer2011}
\bibinfo{author}{\bibfnamefont{P.~W.} \bibnamefont{Brouwer}},
  \bibinfo{author}{\bibfnamefont{M.}~\bibnamefont{Duckheim}},
  \bibinfo{author}{\bibfnamefont{A.}~\bibnamefont{Romito}}, \bibnamefont{and}
  \bibinfo{author}{\bibfnamefont{F.}~\bibnamefont{von Oppen}},
  \bibinfo{journal}{Phys. Rev. B} \textbf{\bibinfo{volume}{84}},
  \bibinfo{pages}{144526} (\bibinfo{year}{2011}),
  \urlprefix\url{http://link.aps.org/doi/10.1103/PhysRevB.84.144526}.

\end{thebibliography}

\end{document}